\documentclass{article}
\usepackage{graphicx}
\usepackage{verbatim}
\usepackage{wrapfig}
\usepackage{color}
\usepackage{cite}
\usepackage{array}
\usepackage{epsfig}
\usepackage{booktabs}
\usepackage{comment}
\usepackage{longtable}
\usepackage{float}
\usepackage{amsmath}
\usepackage{amssymb}
\usepackage{bm}
\usepackage[american]{circuitikz}
\usepackage[latin1]{inputenc}
\usepackage{tikz}
\usetikzlibrary{shapes,arrows,chains,scopes,positioning}
\pdfoutput=1

\newcommand\scalemath[2]{\scalebox{#1}{\mbox{\ensuremath{\displaystyle #2}}}}

\title{Construction of Energy Functions for Lattice Heteropolymer Models: A Case Study in Constraint Satisfaction Programming and Adiabatic Quantum Optimization}
\author{Ryan Babbush$^1$\thanks{Corresponding author. Electronic address: babbush@fas.harvard.edu.}, Alejandro Perdomo-Ortiz$^2$, Bryan O'Gorman$^1$,\\
William Macready$^3$, and Alan Aspuru-Guzik$^1$\thanks{Corresponding author. Electronic address: alan@aspuru.com.}}
\date{\today }

\begin{document}

\maketitle

\begin{abstract}
Optimization problems associated with the interaction of linked particles are at the heart of polymer science, protein folding and other important problems in the physical sciences. In this review we explain how to recast these problems as constraint satisfaction problems such as \textsc{linear programming}, \textsc{maximum satisfiability}, and \textsc{pseudo-boolean optimization}. By encoding problems this way, one can leverage substantial insight and powerful solvers from the computer science community which studies constraint programming for diverse applications such as logistics, scheduling, artificial intelligence, and circuit design. We demonstrate how to constrain and embed lattice heteropolymer problems using several strategies. Each strikes a unique balance between number of constraints, complexity of constraints, and number of variables. Finally, we show how to reduce the locality of couplings in these energy functions so they can be realized as Hamiltonians on existing quantum annealing machines. We intend that this review be used as a case study for encoding related combinatorial optimization problems in a form suitable for adiabatic quantum optimization.
\end{abstract}

\begin{center}
$^1$Department of Chemistry and Chemical Biology, Harvard University,\\ 12 Oxford Street, Cambridge, MA 02138, USA\\
$^2$NASA Ames Quantum Laboratory, Ames Research Center,\\ Moffett Field, CA 94035, USA\\
$^3$D-Wave Systems, Inc., 100-4401 Still Creek Drive, Burnaby,\\ British Columbia V5C 6G9, Canada\\
\end{center}
\newpage

\tableofcontents

\newpage
\section{Introduction}
\subsection{Motivation and Background}
Optimization problems associated with the interaction of linked particles are ubiquitous in the physical sciences. For example, insights into a problem of biological relevance such as the protein folding problem can be obtained from trying to solve the optimization problem of finding the lowest energy configuration of a given sequence of amino acids in space \cite{Sali1994,Pande2010,Dill2008,Mirny2001,Pande2000,Dill1995,Gruebele1998,Shakhnovich1994}. Among other examples of biologically relevant polymers, DNA and RNA chains also fold into complicated structures which can be challenging to predict.

The number of possible configurations (in fact, the number of local minima) for a protein with $N$ amino acids is exponential in $N$ \cite{Hart1997}. Even the simplest model for lattice folding \cite{Lau1989} was proved to be an \textsc{NP-hard} problem \cite{Berger1998,Crescenzi1998}. This implies that the scaling of the worst case scenario for arbitrary protein sequences is exponential with the size of the system. This scaling imposes limitations on the exhaustive search in lattice models for proteins with as few as 36 amino acids in even the most coarse grained protein models \cite{Schram2011}.

An alternative route to exhaustive search or the development of new heuristics is to map these problems into the form of other, more general problems which have been extensively studied for decades. For instance, the \textsc{NP-Hard} problem known as \textsc{Max-SAT} has central importance to practical technologies such as artificial intelligence, circuit design, automated theorem proving, cryptography and electronic verification \cite{Hansen1990,Hansen1998,Soos2009}. The study of this particular problem is central to computer science. There are several journals, conferences and competitions every year dedicated entirely to solving \textsc{SAT} problems \cite{Marques-Silva2007}. Another widely studied constraint satisfaction problem is \textsc{linear programming} which has many applications including logistics scheduling, operations research, company management, and economic planning \cite{Hemmecke2009}. Some applications of linear programming, i.e. multi-commodity flow problems, are considered important enough that entire fields of research exist to develop specialized algorithms for their solution \cite{Even1976}. Once cast as one of these canonical constraint satisfaction problems one can leverage decades of progress in these fields to solve lattice heteropolymer problems. Though it has received relatively little attention until recently, the idea that constraint programming can help solve problems of this type has at least appeared in protein folding and computer science literature since \cite{Yue1995}. Other relevant papers include \cite{Ullah2010,DalPalu2004,Krippahl1999,Backofen1998,Backofen2006}.

Another intriguing option is to study these problems using a computer which takes advantage of quantum mechanical effects to drastically reduce the time required to solve certain problems. For combinatorial optimization problems, perhaps the most intuitive quantum computing paradigms is quantum annealing
\cite{Finnila1994,Kadowaki1998,Santoro2006,Das2008,DeFalco1988,DeFalco1989,Smelyanskiy2012}, a subset of adiabatic quantum optimization \cite{Farhi2000,Farhi2001,Kadowaki1998}. In quantum annealing, the presence of quantum fluctuations (tunneling) allows the system to efficiently traverse potential energy barriers which have a tendency to trap classical optimizations algorithms. Motivated by the experimental realization of studying biologically interesting optimization problems with quantum computation, in this contribution we present a general construction of the free-energy function for the two-dimensional lattice heteropolymer model widely used to study the dynamics of proteins. While the authors have already demonstrated some of these techniques in 
\cite{Perdomo2008}, the encoding strategies discussed here are more general and also more efficient than what we have explained previously. The reduction in resources achieved with these methods allowed for the first experimental implementation of lattice folding on a quantum device \cite{Perdomo-Ortiz2012} where we employed up to 81 superconducting qubits to solve a 6 amino-acid problem in the Miyazawa-Jernigan (MJ) model \cite{Miyazawa1996}.

The goal of this review is to explain the mapping used in \cite{Perdomo-Ortiz2012}, to discuss the strengths and weaknesses of this mapping with respect to other strategies, and to demonstrate how to map the lattice heteropolymer problem into forms which can be solved by using different types of technology and algorithms. While the focus of this paper will be on lattice protein folding, the methods introduced here have very general relevance to discrete and combinatorial optimization problems in science. Whether one decides to use a classical or a quantum (annealing) device, the mappings and techniques presented here emphasize the importance of three key considerations: energy function locality, coupler/coefficient resolution, and efficiency of encoding.

\begin{figure}[H]
\label{flow}
\centering
\begin{tikzpicture}[node distance = 3.2cm, auto]
\tikzset{top/.style={draw, rectangle, rounded corners, thick, text centered, fill=red!20,
text width =20em, minimum height=2.0em}}
\tikzset{cloud/.style={draw, ellipse, thick, text centered, fill=green!20,
text width = 4em, execute at begin node=\small}}
\tikzset{block/.style={draw, rectangle, rounded corners, thick, text centered, fill=red!20,
text width = 8em, minimum height=4em, execute at begin node=\small}}
\tikzset{decision/.style={draw, diamond, thick, text centered, fill=blue!20, inner sep=0pt,
text width = 5.8em, execute at begin node =\small}}
\tikzset{answer/.style={text centered, execute at begin node=\small}}
\tikzstyle{line} = [draw, very thick, -latex']

    \node [top, node distance = 3cm] (init) {What is your computing paradigm?};
    \node [block, left of=init, below of=init] (resourcesQ) {Is number of bits or coupler resolution more limiting?};
    \node [block, right of=init, below of=init] (heuristicQ) {Is the problem small enough to solve exactly?};
    \node [block, below of=resourcesQ] (couplingsQ) {Does your device have programmable many-body terms?};
    \node [block, below of=couplingsQ] (2local) {Does your device have 3-local couplings?};

    \node [cloud, below of=2local] (qubo) {Reduce to QUBO};
    \node [cloud, below of=init] (sat) {Try SAT heuristic};
    \node [cloud, below of=heuristicQ] (ilp) {Try ILP solver};
    \node [cloud, below of=ilp] (pbo) {Try PBO solver};
    \node [cloud, below of=pbo] (sat2) {Try SAT solver};

    \node [decision, below of=resourcesQ, right of=resourcesQ] (diamond) {Diamond construction};
    \node [decision, below of=diamond] (circuit) {Turn circuit construction};
    \node [decision, below of=circuit] (ancilla) {Turn ancilla construction};
    
    \path [line] (init) -- node [answer,right, text width = 7 em]{Traditional Computation}(heuristicQ);
    \path [line] (init) -- node [answer,left, text width = 7 em]{Quantum Annealing}(resourcesQ);
    \path [line] (resourcesQ) -- node [answer]{Couplers}(diamond);
    \path [line] (resourcesQ) -- node [answer, left]{Bits}(couplingsQ);
    \path [line] (couplingsQ) -- node [answer]{Yes}(circuit);
    \path [line] (couplingsQ) -- node [answer, left]{No}(2local);
    \path [line] (2local) -- node [answer]{Yes}(ancilla);
    \path [line] (2local) -- node [answer,left]{No}(qubo);
    \path [line] (qubo) -- (ancilla);

    \path [line] (heuristicQ) -- node[answer,above]{No}(sat);
    \path [line] (sat) -- (diamond);

    \path [line] (heuristicQ) -- node[answer]{Perhaps}(ilp);
    \path [line] (ilp) -- (diamond);
    \path [line] (ilp) -- node[answer]{Solver failed}(pbo);
    \path [line] (pbo) -- (circuit);
    \path [line] (pbo) -- node[answer]{Solver failed}(sat2);
    \path [line] (sat2) -- (ancilla);
   
\end{tikzpicture}
\caption{Flow chart describing how one might choose between the three problem encodings discussed in this review based based on available computing resources. The ``diamond encoding'' is not very efficient but produces a sparse \textsc{QUBO} matrix without requiring reductions that increase the required coupler resolution. This makes it a natural choice for classical integer-linear programming (ILP) and heuristic satisfiability (SAT) solvers which perform best on underconstrained problems. The ``turn circuit'' representation is an overconstrained, but highly efficient, mapping that works best for methods designed to solve high-local expressions such as many-body ion trap simulators or \textsc{pseudo-boolean optimization} (PBO) solvers. The ``turn ancilla'' encoding represents a balance of these benefits as it is relatively efficient and can easily collapse to 2-local without extremely high term coefficients.}
\end{figure}
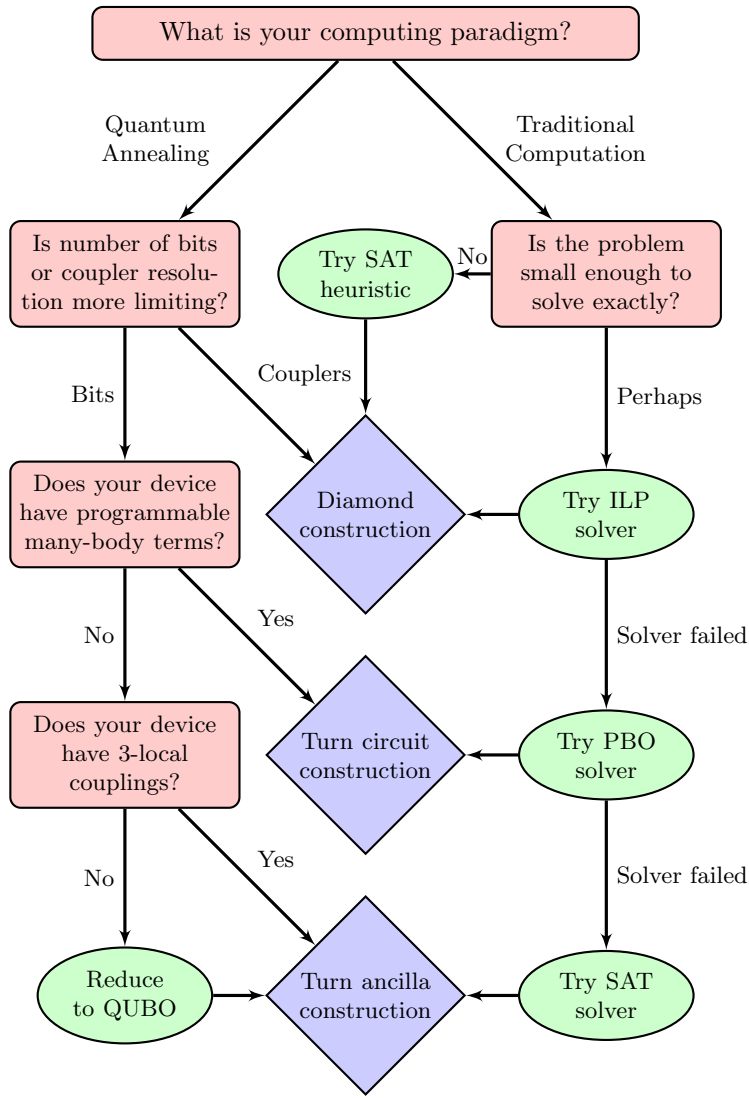

In this context, the ``locality'' of an expression refers to the order of the largest many-body expansion term. For instance, \textsc{quadratic unconstrained binary optimization} (\textsc{QUBO}) problems, which are a binary version of the Ising model, are said to be ``2-local'' because \textsc{QUBO} expressions never contain terms with more than two variables. This is a relevant consideration because an expression which is 3-local cannot be programmed into a quantum device with only pairwise couplings. A similar consideration applies to classical solvers. Coefficient resolution refers to the ability of a quantum device or classical solver to program coupler values to the degree of precision required for the problem. Finally, the efficiency of the encoding refers to the number of bits required to encode the problem. A sketch of how one might weigh these considerations to determine an encoding is shown in Fig.~\ref{flow}.

\subsection{Overview of Mapping Procedure}

The embedding strategies presented here apply to many discrete optimization problems. Mapping these problems to a constraint programming problem is a three step process. In this section we provide a brief description of the process and expand upon each step as it applies to lattice folding in later sections.

\begin{enumerate}
\item \textbf{Encode solution space in computational basis} \\
Define a one-to-one mapping between possible valid assignments of the problem and a bit string encoding this information. Let us denote the bit string by $\boldsymbol{q} \equiv q_1 q_2 \cdots q_n$. The way information is encoded at this point can drastically alter the nature of the following three steps so one must take care to choose a mapping which will ultimately make the best use of resources; in many cases, the most compact mapping will have a high order energy function or require many ancillary bits. Regardless of how information is encoded, the bit string must uniquely enumerate each element of the low energy solution space.

\item \textbf{Constrain energy landscape with pseudo-boolean expression} \\
Construct a pseudo-boolean energy function $E(\boldsymbol{q}) = E(q_1, q_2, \cdots, q_n)$ which takes $\boldsymbol{q}$ as input and correctly reproduces the relative energies in the low energy subspace of the original problem so that the optimal solution to $E(\boldsymbol{q})$ encodes the solution to the original problem. The construction of this function is not trivial and will depend largely on how information is encoded in $\boldsymbol{q}$. At this point it may be necessary to increase the dimensionality of the solution space by adding ancillary bits. In a previous contribution, we provided a specific technique to construct the energy function for particles interacting in a lattice \cite{Perdomo2008}. The purpose of this contribution is to introduce the reader to several different types of mappings which have distinct advantages or disadvantages depending on problem size, complexity and available resources.

\item \textbf{Map boolean representation to desired constraint programming} \\
In most cases one can take advantage of significantly more powerful solvers by making a final transformation from pseudo-boolean function to \textsc{weighted maximum satisfiability} (W-SAT), \textsc{integer-linear programming} (ILP), or QUBO. When cast as a W-SAT problem, one can take advantage of both heuristic and exact W-SAT solvers which have been developed by the computer science community and tested every year in annual ``SAT Competitions". When represented as an ILP problem, one can use commercial logistics scheduling software such as IBM's CPLEX. If one wishes to implement the energy expression on a quantum device it may be necessary to manipulate the energy expression so that it contains only local fields and two-body couplings. Thus, the final step is often to reduce the dimensionality of the pseudo-boolean expression to 2-local so that the problem can be implemented as QUBO on currently existing architectures for adiabatic quantum computing as was done in \cite{Perdomo-Ortiz2012}.
\end{enumerate}

\newpage
\section{The ``Turn" Encoding of Self-Avoiding Walks}
\subsection{Embedding physical structure}

Let us use the term ``fold" to denote a particular self-avoiding walk (SAW) assumed by the ordered chain of beads or ``amino acids'' on a square lattice. These configurations include amino acid chains that might intersect at different points due to amino acids occupying the same lattice sites. Even though overlapping folds will exist in the solution space of our problem, these folds are unphysical and therefore we need to construct energy functions to penalize such configurations. Such functions will be discussed in detail below.

A fold of an $N$ amino acid protein is represented in what we refer to as the ``turn" mapping by a series of $N-1$ turns. We use this name to distinguish the encoding from other (spatial) representations which encode the possible folds by explicitly encoding the grid location of each amino acid. The square lattice spatial representation discussed in \cite{Perdomo2008} has the advantage of being general for the problem of $N$ particles interacting in a lattice (which need not be connected) but we can do much better in terms of the number of variables needed; bit efficiency is the main advantage of the turn mapping.

In the turn mapping, one saves bits by taking advantage of the connectivity of a valid SAW to store information about where each amino acid is relative to the previous amino acid instead of encoding explicit amino acid locations. Therefore, instead of encoding the positions of the $j$th amino acids in the lattice, we encode the $j$th turn taken by the $j+1$ amino acid in the chain. For pedagogical purposes, we concentrate on the case of a two-dimensional $(2D)$ lattice SAW; the extension to a three-dimensional lattice requires a straightforward extension of the same techniques described here for the $2D$ case.

\begin{figure}[h]
 \centering
  \includegraphics[scale=0.5]{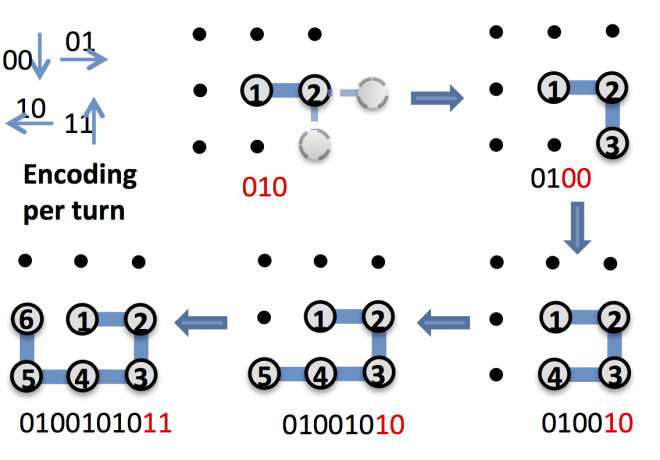}
 \caption{Step-by-step construction of the binary representation of a particular six unit lattice protein in the turn encoding. Two qubits per bond are needed and the turn ``compass'' (bond directions) are denoted as ``00'' (downwards), ``01'' (rightwards), ``10'' (left), and ``11'' (upwards). This image has been reproduced from \protect\cite{Perdomo-Ortiz2012} with permission from the authors.}
  \label{turns}
\end{figure}

Because the location of an amino acid in the turn mapping is specified by its location relative to the previous acid in the primary sequence, the solution space consists only of paths, or ``worms'', embedded in the lattice. The resulting energy function is invariant under translational, rotation and reflection with respect to the embedding in physical space as long as the local structure of the relative locations is kept intact. More specifically, each of the $N-1$ turns in $2D$ space requires two bits so that each of the four directions (up, down, left, and right) has a unique representation. This assumes a rectilinear lattice, but the method is equally valid, though with slight modification, for other lattices, e.g. triangular. The convention or ``compass'' used in this paper is presented in the upper-left part of Fig.~\ref{turns}. Furthermore, we can fix the first three bits to obtain only solutions which are rotationally invariant. Under this convention, the bit-string $\boldsymbol{q}$ is written as,
\begin{equation}
\boldsymbol{q} = 0 1 \underbrace{0 q_1}_{turn2} \underbrace{q_2 q_3}_{turn3} \cdots \underbrace{q_{2(N-1)-4} q_{2(N-1)-3}}_{turn(N-1)}
\label{bits}
\end{equation}

We have chosen to fix the first three bits as 010 so that the walk always turns first to the right and then either right or down. This does not  affect the structure of the solution space and leaves only $N-2$ turns to be specified; an example is provided in Eq.~\ref{bits}. Since every turn requires 2 bits, the turn mapping requires only $2 (N-2)-1 = 2N-5$ bits to represent a fold. This can be compared with the $(2N-4)\log_2N$ required for the spatial mapping in \cite{Perdomo2008}. To clearly demonstrate how this mapping works, an example of the turn encoding for a short SAW is shown below in Fig.~\ref{turns}.

\subsection{``Turn ancilla'' construction of $E(\boldsymbol{q})$}

Now that we have a mapping function which translates a length $2N - 5$ bit-string into a specific fold in the 2D lattice we can construct $E(\boldsymbol{q})$ as a function of these binary variables. For the case of lattice folding, we need to penalize folds where two amino acids overlap, i.e. the chain must be self-avoiding. This penalty will be given by the energy function, $E_{overlap}\left(\boldsymbol{q}\right)$, which returns an extremely high value if and only if amino acids overlap. While it is possible to construct a single function $E_{overlap}\left(\boldsymbol{q}\right)$ which penalizes all potential overlaps, we will show that less ancillary bits are needed if we introduce the function $E_{back}\left(\boldsymbol{q}\right)$ which penalizes the special case of overlaps that happen because the chain went directly backwards on itself. In this scheme, $E_{overlap}\left(\boldsymbol{q}\right)$ will apply to all other potential overlaps.

Finally, we must consider the interaction energy among the different amino acids. This will ultimately determine the structure of the lowest energy fold. The energy given by the pairwise interaction of beads in our chain will be given by $E_{pair}(\boldsymbol{q})$. In some lattice protein models such as the Hydrophobic-Polar (HP) protein folding model, there is only one stabilizing interaction; however, the construction we present here applies for an arbitrary interaction matrix among the different amino acids (or particles to be even more general). One of the advantages of the turn representation over the spatial representation is that we do not need to worry about having the amino acids linked in the right order (primary sequence), since this is guaranteed by design. The construction of the energy function,
\begin{equation}\label{eq:Etotal}
E(\boldsymbol{q}) = E_{back}\left(\boldsymbol{q}\right) + E_{overlap}(\boldsymbol{q}) + E_{pair}(\boldsymbol{q}),
\end{equation}
involves a series of intermediate steps which we outline next.

\subsubsection{Construction of $E_{back}(\boldsymbol{q})$}
In order to have a valid SAW we need to guarantee that our ``worm'' does not turn left and then immediately turn right or vice versa or turn up and then immediately turn down or vice versa. In order to program this constraint into the energy function we will introduce several simple logic circuits. Looking at the compass provided in Fig.~\ref{turns} it should be clear the circuits in Figs.~\ref{circuits1}-\ref{circuits4} return {\sc true} if and only if a particular turn (encoded $q_{1} q_{2}$) went right, left, up, or down respectively.
\begin{figure}[H]
\small
\caption{A logical circuit representing ``right'' consisting of a {\sc not} gate after the first bit and an {\sc and} gate. Evaluates to {\sc true} if and only if $q_1,q_2 = 0,1$.}
\centering
\tikzstyle{branch}=[fill,shape=circle,minimum size=5pt,inner sep=0pt]
\begin{circuitikz}
\draw
node (a) at (0,2) {$q_{1}$}
node (b) at (1,2) {$q_{2}$}

node[scale=0.6, not port, rotate=-90] (nota) at (0,1) {}
node[and port] (and1) at (3, 0) {}

(a) |- (nota.in)
(nota.out) |- (and1.in 1)
(b) |- (and1.in 2)
(and1.out) node[right] {$\left(q_{2} - q_{1} q_{2}\right)$};
\end{circuitikz}
\label{circuits1}

\centering
\caption{A logical circuit representing ``left'' consisting of a {\sc not} gate after the second bit and an {\sc and} gate. Evaluates to {\sc true} if and only if $q_1,q_2 = 1,0$.}
\begin{circuitikz}
\draw
node (a) at (0,2) {$q_{1}$}
node (b) at (1,2) {$q_{2}$}

node[scale=0.6, not port, rotate=-90] (notb) at (1,1) {}
node[and port] (and1) at (3, 0) {}

(a) |- (and1.in 1)
(b) |- (notb.in)
(notb.out) |- (and1.in 2)
(and1.out) node[right] {$\left(q_{1} - q_{1} q_{2}\right)$};
\end{circuitikz}
\label{circuits2}
\end{figure}

\begin{figure}[H]
\centering
\caption{A logical circuit representing ``up''. Only {\sc true} if $q_1,q_2 = 1,1$.}
\begin{circuitikz}
\draw
node (a) at (0,1) {$q_{1}$}
node (b) at (1,1) {$q_{2}$}

node[and port] (and1) at (3,0) {}
(a) |- (and1.in 1)
(b) |- (and1.in 2)
(and1.out) node[right] {$\left(q_{1} q_{2}\right)$};
\end{circuitikz}
\label{circuits3}

\centering
\caption{A logical circuit representing ``down''. Only {\sc true} if $q_1,q_2 = 0,0$.}
\begin{circuitikz}
\draw
node (a) at (0,2) {$q_{1}$}
node (b) at (1,2) {$q_{2}$}

node[scale=0.6, not port, rotate=-90] (nota) at (0,1) {}
node[scale=0.6, not port, rotate=-90] (notb) at (1,1) {}
node[and port] (and1) at (3, 0) {}

(a) |- (nota.in)
(nota.out) |- (and1.in 1)
(b) |- (notb.in)
(notb.out) |- (and1.in 2)
(and1.out) node[right] {$\left(1 - q_{1} - q_{2} + q_{1} q_{2}\right)$};
\end{circuitikz}
\label{circuits4}
\end{figure}
\normalsize

\begin{figure}[h]
\centering
\tikzstyle{branch}=[fill,shape=circle,minimum size=5pt,inner sep=0pt]
\begin{circuitikz}
\draw
node (a) at (0,4) {$q_{i}$}
node (b) at (0.75,4) {$q_{i+1}$}
node (c) at (1.5,4) {$q_{i+2}$}
node (d) at (2.25,4) {$q_{i+3}$}

node[scale=0.6, not port, rotate=-90] (nota) at (0,3) {}
node[scale=0.6, not port, rotate=-90] (notd) at (2.25,3) {}

node[and port] (and1) at (4,2) {}
node[and port] (and3) at (6,1) {}
(a) |- (nota.in)
(nota.out) |- (and1.in 1)
(b) |- (and1.in 2)

(and1.out) |- (and3.in 1)

node[and port] (and2) at (4,0) {}

(c) |- (and2.in 1)
(d) |- (notd.in)
(notd.out) |- (and2.in 2)
(and2.out) |- (and3.in 2)
(and3.out) node[right] {$\left(d^j_{x+} \wedge d^{j+1}_{x-}\right)$};
\end{circuitikz}
\caption{A logical circuit which returns {\sc true} if and only if $\left(d^j_{x+} \wedge d^{j+1}_{x-}\right)$, i.e. the turn sequence $q_i q_{i+1} q_{i+2} q_{i+3} = 0110$, meaning it went right and then left.}
\label{back1}
\end{figure}
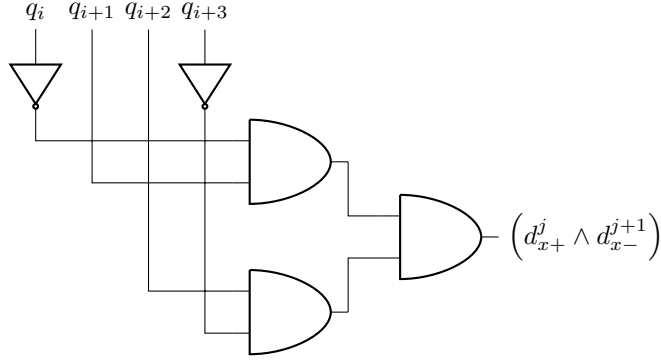

Using these circuits we can generalize the concept of ``up'', ``down'', ``left'' and ``right'' functions to precise directional strings. In two dimensions (as prescribed by Fig.~\ref{turns}), we have the functions for the $j$th turn,
\begin{eqnarray}
d^j_{x+}&=&q_{2j-3}(1-q_{2j-4})=q_{2j-3}-q_{2j-3}q_{2j-4}\\
d^j_{x-}&=&(1-q_{2j-3})q_{2j-4}=q_{2j-4}-q_{2j-3}q_{2j-4}\\
d^j_{y+}&=&q_{2j-3}q_{2j-4}\\
d^j_{y-}&=&(1-q_{2j-3})(1-q_{2j-4})=1-q_{2j-3}-q_{2j-4}+q_{2j-3}q_{2j-4},
\label{directional}
\end{eqnarray}
which evaluate to {\sc true} if and only if the $j$th turn is to be right, left, up or down respectively. Having defined these circuits we can construct a more complicated circuit which takes two turns (the 4 bits $q_{i}q_{i+1}q_{i+2}q_{i+3}$) as input and returns {\sc true} if and only if the second turn went backwards, i.e. $\left(d^j_{x+} \wedge d^{j+1}_{x-}\right)\vee\left(d^j_{x-}\wedge d^{j+1}_{x+}\right)\vee\left(d^j_{y+}\wedge d^{j+1}_{y-}\right) \vee \left(d^j_{y-} \wedge d^{j+1}_{y+}\right)$. An example of these conjunctions, $\left(d^j_{x+} \wedge d^{j+1}_{x-}\right)$ is shown in Fig.~\ref{back1}.
\small
\normalsize

The other three conjunctions are also trivially constructed by combining the appropriate circuits using {\sc and} gates which simply multiply together the directional strings. The utility of these circuits is that they produce terms in a pseudo-boolean function. Specifically we get the terms,
\small
\begin{eqnarray}
\label{def1}
\left(d^j_{x+} \wedge d^{j+1}_{x-}\right) &=& q_{i+1}q_{i+2} - q_{i}q_{i+1}q_{i+2} - q_{i+1}q_{i+2}q_{i+3} + q_{i}q_{i+1}q_{i+2}q_{i+3}\\
\label{def2}
\left(d^j_{x-} \wedge d^{j+1}_{x+}\right) &=& q_{i}q_{i+3}-q_{i}q_{i+1}q_{i+3}-q_{i}q_{i+2}q_{i+3}+q_{i}q_{i+1}q_{i+2}q_{i+3}\\
\label{def3}
\left(d^j_{y+} \wedge d^{j+1}_{y-}\right) &=& q_{i}q_{i+1}-q_{i}q_{i+1}q_{i+2}-q_{i}q_{i+1}q_{i+3}+q_{i}q_{i+1}q_{i+2}q_{i+3}\\
\label{def4}
\left(d^j_{y-} \wedge d^{j+1}_{y+}\right) &=& q_{i+2}q_{i+3}-q_{i}q_{i+2}q_{i+3}-q_{i+1}q_{i+2}q_{i+3}+q_{i}q_{i+1}q_{i+2}q_{i+3},
\end{eqnarray}
\normalsize
where $i=2j-4$.
It might seem logical to finish this circuit by combining all four backwards overlap circuits with {\sc or} gates; however, this is not an advisable strategy as it is sure to produce many high ordered terms. Because exactly one or none of these circuits will be {\sc true} we can accomplish the same result by summing the four circuits. Accordingly, for the two turns $q_{i}q_{i+1}q_{i+2}q_{i+3}$ the pseudo-boolean expression,
\begin{equation}
\left(d^j_{x+} \wedge d^{j+1}_{x-}\right) + \left(d^j_{x-} \wedge d^{j+1}_{x+}\right) + \left(d^j_{y+} \wedge d^{j+1}_{y-}\right) + \left(d^j_{y-} \wedge d^{j+1}_{y+}\right)
\label{psuedo-back}
\end{equation}
evaluates to {\sc true} if and only if $q_{i}q_{i+1}q_{i+2}q_{i+3}$ represents a backwards turn and evaluates to {\sc false} otherwise. Our goal is to construct a pseudo-boolean expression which returns a penalty whenever a backwards turn is made; therefore we must multiply this expression by a constant to be determined later known as $\lambda_{overlap}$. After substituting Eqs.~\ref{def1}-\ref{def4} into Eq.~\ref{psuedo-back}, factoring the terms, and adding in $\lambda_{overlap}$ we can write,
\small
\begin{equation}
E_{back}\left(q_{i}q_{i+1}q_{i+2}q_{i+3}\right) = \lambda_{overlap} \left(2q_{i}q_{i+2}-q_{i}-q_{i+2}\right)\left(2q_{i+1}q_{i+3}-q_{i+1}-q_{i+3}\right).
\end{equation}
\normalsize

To construct the entire $E_{back}\left(\boldsymbol{q}\right)$ we need to sum together bits from each pair of adjacent turns. Keeping in mind that we fix the first three bits at 010, we write the final expression for $E_{back}(\boldsymbol{q})$ as,
\small
\begin{equation}
\begin{array}{rcl}
E_{back}(\boldsymbol{q}) & = & \lambda_{overlap}\left(q_{1}q_{2}+q_{2}q_{3}-2q_{1}q_{2}q_{3}\right)\\
& + & \lambda_{overlap} \sum_{i=2}^{2 N-8}E_{back}\left(q_{i}q_{i+1}q_{i+2}q_{i+3}\right)\left[(i+1)\textrm{mod}2\right].
\end{array}
\label{eback}
\end{equation}
\normalsize
In this expression, the first three terms come from ensuring that the second turn (which begins with a bit fixed at 0) does not overlap with the third turn. Notice that in this expression, the first physical bit with an unknown value is labeled ``$q_{1}$'' despite the fact that the first three information bits are fixed at 010. This formalism will be consistent throughout our review.

It is important to point out that while the decision to use a separate $E_{back}(\boldsymbol{q})$ instead of a more general $E_{overlap}\left(\boldsymbol{q}\right)$ has the disadvantage of introducing 3 and 4-local terms, it has the advantage of construction without any ancillary bits. Furthermore, even if one needs an entirely 2-local expression this strategy may still be preferable because the same reductions needed to collapse this expression to 2-local will be needed in collapsing the pairwise energy function to 2-local by construction. For more on reductions, see Sec.~\ref{reduction}.

\subsubsection{Construction of $E_{overlap}\left(\boldsymbol{q}\right)$ with ancilla variables}
The overlap energy function $E_{overlap}(\boldsymbol{q})$ penalizes configurations in which any two amino acids share the same lattice point. The penalty energy associated with any pair of amino acids overlapping must be large enough to guarantee that it does not interfere with the spectrum of the valid configurations (we return to the topic of choosing penalty values later on). We begin by defining a function which specifies the $x$ and $y$ grid positions of each amino acid. Because the directional strings we defined earlier in Eqs.~\ref{def1}-\ref{def4} keep track of the direction of every step we can define these functions as
\begin{eqnarray}
x_{n} &=& 1+q_{1}+\sum_{k=3}^{n-1}\left(d^k_{x+} - d^k_{x-}\right) \text{ and}\\
y_{n} &=& q_{1}-1+ \sum_{k=3}^{n-1}\left(d^k_{y+} - d^k_{y-}\right),
\end{eqnarray}
where the position of the $n$th amino acid in the sequence is a function of the preceding $n-1$ turns iterated through with index $k$. Note that the terms in front of the sum are determined by the first three (fixed) bits: 010. With these definitions we can make an extremely useful function which will return the square of the grid distance between any two amino acids (denoted $i$ and $j$):
\begin{equation}
g_{ij} = \left(x_{i}-x_{j}\right)^2 + \left(y_{i}-y_{j}\right)^2.
\label{g}
\end{equation}
$g_{ij}$ has several extremely useful properties worth pointing out now. First, $g_{ij}$ is zero if and only if two amino acids overlap; otherwise, $g_{ij}$ is always positive. Additionally, $g_{ij}$ has the very surprising property of being natively 2-local when constructed using the compass that we defined in Fig.~\ref{turns} (therefore the decision to encode directions in that fashion was not arbitrary). This is surprising because the directional strings are 2-local so we might na\"\i vely expect something which involves the square of these to be 4-local; however this turns out not to be the case because $x_{n}$ and $y_{n}$ are 1-local by construction.

In order to use $g_{ij}$ to construct $E_{overlap}\left(\boldsymbol{q}\right)$ we need a function which takes $g_{ij}$ as input and returns a penalty if and only if $g_{ij} = 0$. First, we note the bounds on $g_{ij}$,
\begin{equation}
0 \leq g_{ij} \leq (i-j)^2.
\end{equation}
To help enforce the constraint that $g_{ij} \geq 1$, we introduce a free parameter, $\alpha_{ij}$. In the optimization literature, such a variable is called a ``slack variable'' and is used to convert an inequality into an equality. In our case,
\begin{equation}
0 \leq \alpha_{ij} \leq (i-j)^2-1
\label{alpha1}
\end{equation}
This implies that,
\begin{equation}
\label{alpha2}
\forall \, g_{ij} \geq 1 \, \exists \, \alpha_{ij} : (i-j)^2 - g_{ij} - \alpha_{ij} = 0.
\end{equation}
Furthermore, if and only if $g_{ij} = 0$,
\begin{equation}
(i-j)^2 - g_{ij} - \alpha_{ij} \geq 1 \, \forall \, \alpha_{ij}.
\label{alpha3}
\end{equation}

In order to introduce a slack variable such as $\alpha_{ij}$ into the construction of our pseudo-boolean function we must encode it using ancilla bits. Ancilla bits are real, unconstrained bits used in the calculation which have no physical significance to the particular problem mapping (i.e. ancilla bits do not tell us anything about a particular protein fold). In using ancilla we increase the dimensionality of the solution space of our problem by introducing extra variables but gain the ability to use those bits in our energy function.

Every pair of amino acids which could possibly overlap will need unique bits to form an $\alpha$ for use in the $E_{overlap}\left(\boldsymbol{q}\right)$ term corresponding to that pair. Only amino acids which are an even number of turns apart can possibly overlap and we are already preventing amino acids which are two turns apart from overlapping with $E_{back}(\boldsymbol{q})$; thus, the number of amino acid pairs which require a slack variable is calculated as,
\begin{equation}
\sum_{i=1}^{N-4} \sum_{j=i+4}^{N} \left[\left(1+i-j\right)\textrm{mod}\,2\right].
\end{equation}
Each $\alpha_{ij}$ can be represented in binary using the corresponding ancilla bits. Using Eq.~\ref{alpha1} we see that the $\alpha_{ij}$ corresponding to amino acid pair $i,j$ can be represented in $\mu_{ij}$ ancilla bits where,
\begin{equation}
\mu_{ij} = \lceil 2\log_{2}\left(i-j\right) \rceil \left[\left(1+i-j\right)\textrm{mod}\,2\right].
\label{mu}
\end{equation}
Therefore, the total number of ancilla bits required to form $E_{overlap}\left(\boldsymbol{q}\right)$ is
\begin{equation}
\sum_{i=1}^{N-4} \sum_{j=i+4}^{N} \mu_{ij}.
\end{equation}
Finally, we can write the formula for a given $\alpha_{ij}$ as
\begin{equation}
\alpha_{ij} = \sum_{k = 0}^{\mu_{ij}-1} q_{c_{ij}+k}2^{k},
\label{exponential}
\end{equation}
where $c_{ij}$ denotes a pointer to the first ancilla bit corresponding to a particular amino acid pair. For instance, if the $E_{overlap}\left(\boldsymbol{q}\right)$ ancilla are in sequential order from lowest index pair to highest index pair and come immediately after the information bits then we could write,
\begin{equation}
c_{ij} = \sum_{m=1}^{i}\left( \sum_{n=m+4}^{N} \mu_{mn}\right) - \sum_{n=j}^{N}\mu_{in}.
\label{pointer}
\end{equation}

However, there are still several problems we must address before we can construct $E_{overlap}\left(\boldsymbol{q}\right)$. To begin with, we originally wanted an $\alpha_{ij}$ which was specifically restricted to the domain given in Eq.~\ref{alpha1} but since we cannot constrain the physical bits in any fashion, Eq.~\ref{mu} and Eq.~\ref{exponential} suggest that our slack variable is actually in the domain given by,
\begin{equation}
0 \leq \alpha_{ij} \leq 2^{\mu_{ij}} - 1.
\label{dom_alpha}
\end{equation}
We should adjust Eq.~\ref{alpha2} and Eq.~\ref{alpha3} so that,
\begin{equation}
\label{alpha4}
\forall \, g_{ij} \geq 1 \, \exists \, \alpha_{ij} : 2^{\mu_{ij}} - g_{ij} - \alpha_{ij} = 0.
\end{equation}
Furthermore, if and only if $g_{ij} = 0$,
\begin{equation}
2^{\mu_{ij}} - g_{ij} - \alpha_{ij} \geq 1 \, \forall \, \alpha_{ij}.
\label{alpha5}
\end{equation}

Finally, there is the question of how to guarantee that $\alpha_{ij}$ is the particular $\alpha_{ij}$ that gives $0$ in Eq.~\ref{alpha4} whenever $g_{ij} \geq 1 $. Even though there exist $\alpha_{ij}$ such that Eq.~\ref{alpha4} evaluates to 0, it is also possible to have $\alpha_{ij}$ such that Eq.~\ref{alpha4} evaluates to a negative value. Negative values would incentivize overlaps instead of penalizing them so to ensure that the lowest energy solution always has $E_{overlap}\left(\boldsymbol{q}\right) = 0$ we square the expression to obtain the following formula,
\begin{equation}
\gamma_{i j} = \lambda_{overlap} \left[2^{\mu_{ij}} - g_{ij} - \alpha_{ij}\right]^{2}.
\label{overlap}
\end{equation}
The expression $\gamma_{ij}$ is effective for our purposes because $\alpha_{ij}$'s restricted domain given by Eq.~\ref{dom_alpha}, promises that $\gamma_{ij}$ can only equal zero if $g_{ij} \geq 1$. $\gamma_{ij}$ is zero only if $g_{ij} \geq 1 \wedge \alpha_{ij} = 2^{\mu_{ij}}-g_{ij}$; thus, the goal is to make $\lambda_{overlap}$ a sufficiently large penalty that all low energy solutions must have no overlaps, i.e. $g_{ij} \geq 1$ for all $ij$, and $\alpha_{ij} = 2^{\mu_{ij}}-g_{ij}$. Finally we can write the final expression,
\begin{equation}
E_{overlap}\left(\boldsymbol{q}\right) = \sum_{i=1}^{N-4}\sum_{j=i+4}^{N}\left[\left(1+i-j\right)\textrm{mod}\,2\right]\gamma_{ij}.
\label{overlapeq}
\end{equation}

Again, we include the term $\left[\left(1+i-j\right)\textrm{mod}\,2\right]$ because only amino acids that are an even number apart have the possibility of overlapping. Furthermore, because overlaps between adjacent amino acids are impossible and overlaps between amino acids two apart are prevented by $E_{back}\left(\boldsymbol{q}\right)$, we start the second sum at $j=i+4$ Accordingly, one should only create ancillary bits for pairs in which $\left(i-j\right)\textrm{mod}\,2 = 0 \wedge i-j \geq 4$. It should now be clear that the reason we introduced $E_{back}\left(\boldsymbol{q}\right)$ was so that we used fewer ancillary bits in this step.

\subsubsection{Construction of $E_{pair}(\boldsymbol{q})$ with ancilla variables}
Finally, we need to construct the pairwise interaction energy function. To do this we need to make an interaction matrix, $J$, which contains all of the pairwise interactions which lower the energy when two amino acids are adjacent on the lattice (thus all elements of $J$ are negative or zero). Note that this interaction matrix must contain many zero-valued elements as many amino acid pairs cannot possibly be adjacent. For instance, only amino acids which are at least three turns apart and an odd number of turns apart can ever be adjacent. Furthermore, depending on the interaction model many of these amino acids might not ``interact''; for instance, in the HP-model only H-H pairs can interact where as in the Miyazawa-Jernigan model all amino acids can interact.

For each potential interaction, we must introduce one ancillary bit denoted $\omega_{ij}$ where $i$ and $j$ denote the amino acids involved in the interaction. $\omega_{ij}$ is essentially a switch which is only ``on'' without incurring an energy penalty if two amino acids are interacting (that is, if $g_{ij} = 1$). We can now write the pairwise interaction term:
\begin{equation}
\varphi_{ij} = \omega_{ij} J_{ij} \left(2-g_{ij}\right)
\end{equation} 

This simple function does everything we need to write the pair function. Because $E_{overlap}\left(\boldsymbol{q}\right)$ ensures that $g_{ij} \geq 1$, we see that $\varphi_{ij}$ is only positive if both $J_{ij}$ and $\omega_{ij}$ are non-zero and $g_{ij}$ is greater than 2. Such solutions will never be part of the low-energy landscape for our problem because the energy could be made lower by trivially flipping the $\omega_{ij}$ ancillary bit. On the other hand, $\varphi_{ij} = J_{ij}$ if and only if $g_{ij} = 1 \wedge \omega_{ij}=1$ which means that the pair is adjacent! Thus, the final form of $E_{pair}(\boldsymbol{q})$ is,
\begin{equation}
E_{pair}(\boldsymbol{q}) = \sum_{i=1}^{N-3} \sum_{j=i+3}^{N} \omega_{ij} J_{ij} \left(2-g_{ij}\right).
\label{paireq}
\end{equation}

\subsection{``Turn circuit'' construction of $E(\boldsymbol{q})$}
The turn ancilla construction has the advantage of providing an energy expression with relatively few many-body terms but it does so at the cost of introducing ancilla bits. If one intends to use a pseudo-boolean solver or a quantum device with adjustable many-body couplings, bit efficiency is much more important than the particular structure of the energy expression. This section explains the so-called ``circuit'' construction which provides optimal efficiency at the cost of introducing high ordered many-body terms. The turn circuit construction (along with reductions explained in Sec.~\ref{reduction}) was used to encode problems into a quantum annealing machine in \cite{Perdomo-Ortiz2012}.

\subsubsection{Sum strings}
The circuit construction works by keeping track of the turns in between amino acids to determine if the amino acids overlap or not. To do this we keep track of the turns in every direction using the directional strings defined in Eqs.~\ref{def1}-\ref{def4}. Using these directional strings we introduce ancillary bits referred to as ``sum strings''. Sum strings are strings of $\lceil\log_{2}(j-i)\rceil$ bits for each segment of the chain between amino acids $i$ and $j$, with $1\leq i < j\leq N$ and $i+1<j$. As in the case of the directional strings, we require one `sum string'' per direction per pair of amino acids to be compared. Each represents, in binary, the number of total turns in a particular direction within the segment.

\begin{figure}[h]
  \centering
    \includegraphics{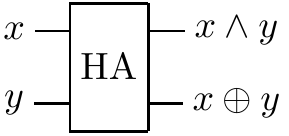}
  \caption{The Half-Adder gate sums two bits.}
\label{ha}
\end{figure}

\begin{figure}[h]
 \centering
  \includegraphics[scale=0.7]{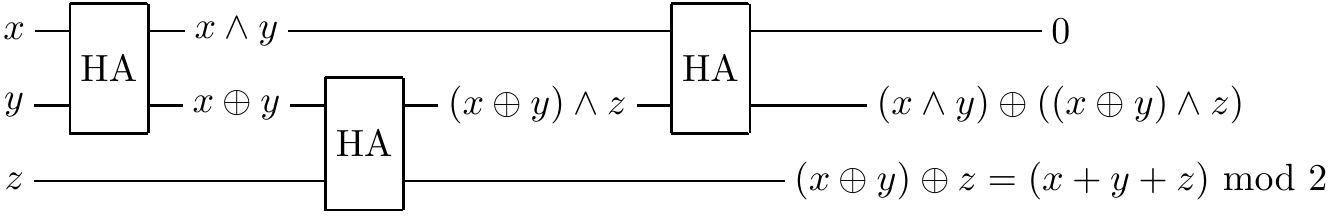}
 \caption{A circuit to sum three bits. }
 \label{sum3}
\end{figure}

As in the ancilla construction, whether or not two amino acids interact or overlap depends on the sequence of turns between them. To determine this, for each segment of the directional strings we construct a string that is the sum, in binary, of the bits between two amino acids, i.e. the total number of turns in that direction. This process is most straightforwardly described using a circuit model. Consider, a single Half-Adder gate (HA) consisting of an {\sc and} and an {\sc xor} gate, as shown in Fig.~\ref{ha}.
The output of a Half-Adder can be interpreted as the two-bit sum of its two input bits. Accordingly, if we wanted to add three bits we could add two of them, and then add the resultant two-bit number to the third bit, as shown in Figure \ref{sum3}.

\begin{figure}[H]
 \centering
  \includegraphics{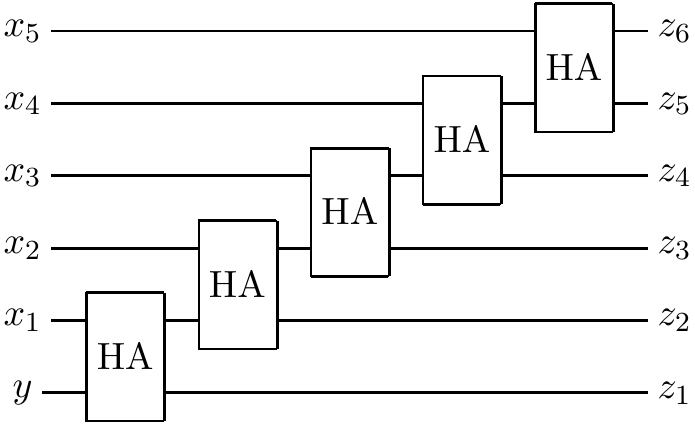}
 \caption{Circuit for the addition of a single bit $y$ to the 5-bit $x=x_5x_4x_3x_2x_1$ to form the 6-bit sum $x+y=z_6z_5z_4z_3z_2z_1$.}
 \label{ripple}
\end{figure}

\begin{figure}[h]
 \centering
  \includegraphics{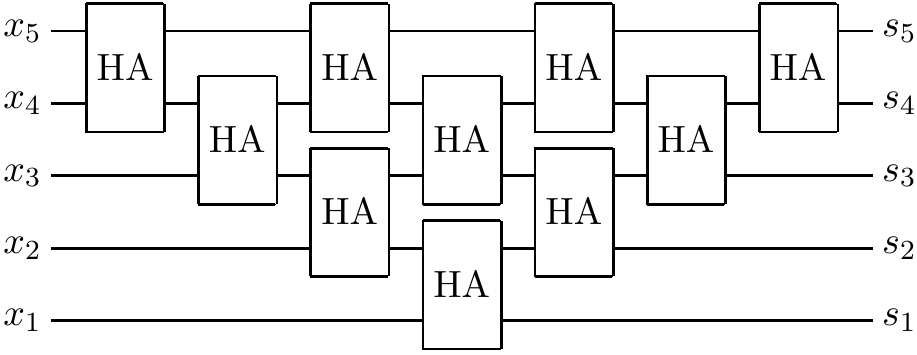}
 \caption{The circuit for the sum, $s_1s_2s_3s_4s_5$, of 5 bits $x_1+x_2+x_3+x_4+x_5$.}
 \label{sum5}
\end{figure}

In general, to add a single bit to an $n$-bit number, we simply apply $n$ Half-Adders. First, a Half-Adder applied to the single bit and the least significant bit of the augend gives the least significant bit of the sum. Next, we use a second Half-Adder to add the carry bit of the first addition and the second least significant bit of the augend to give the second least significant bit of the sum. This process is repeated until the $(n+1)$-bit sum is computed. For an example of this see Fig.~\ref{ripple}.
Thus, given an arbitrary number of bits we can find their sum, in binary, by successively combining the strategies shown in Fig.~\ref{sum5}, i.e. first adding the first three bits (see the first three HA gates from left to right) and then adding the next bit to the resulting three bit number which carries the previous sum. This is accomplished by the next three HA gates. From then and on, one adds a simple bit to each of the resulting $n$ bit number by using $n$ HA gates until all bits in the string are added.
\begin{figure}[H]
 \centering
  \includegraphics[scale=0.7]{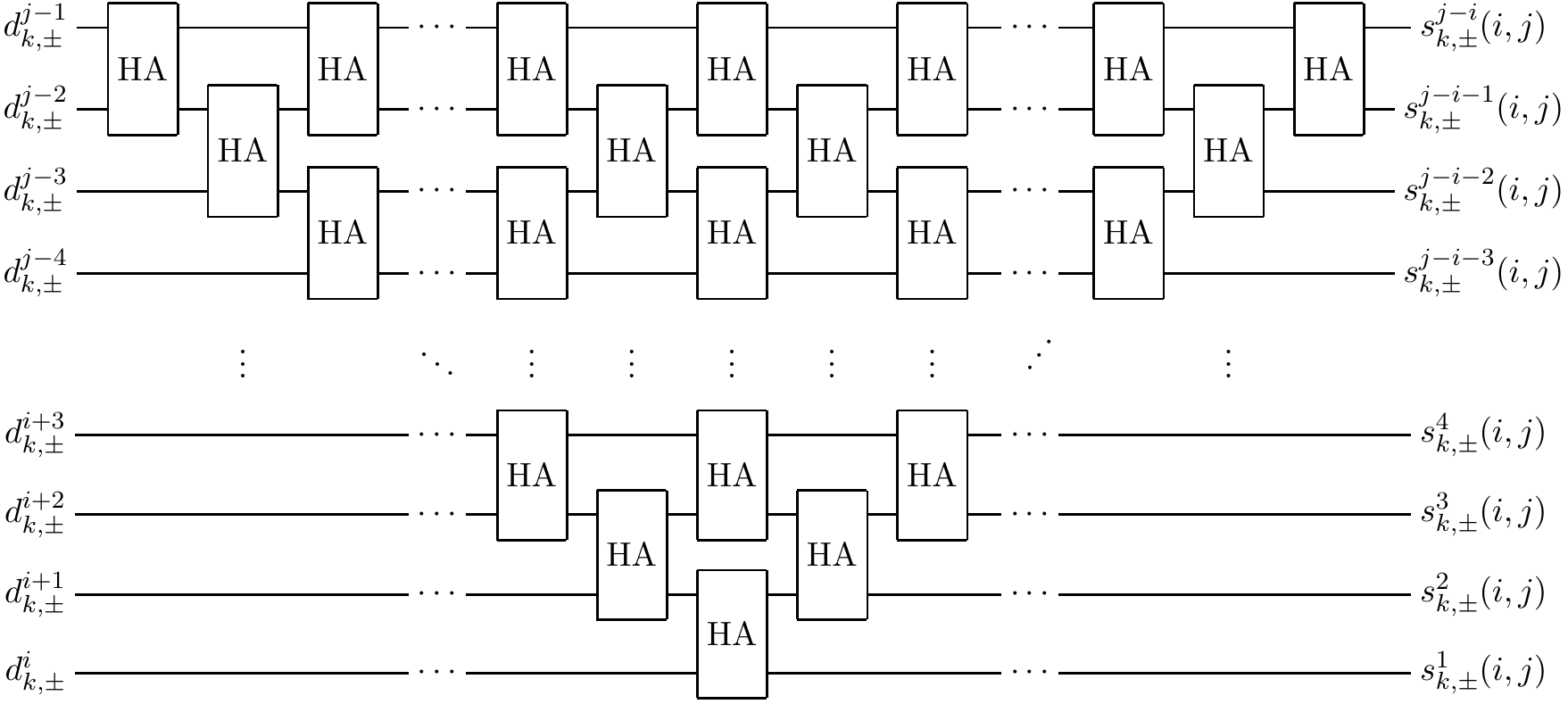} 
\caption{The circuit for the number $s_{k\pm}(i,j)$ of turns between amino acids $i$ and $j$ in the $\pm k$ direction.}
  \label{genSum}
\end{figure}

We can use the circuit such shown in Fig.~\ref{genSum} to compute the the binary digits of a particular sum that will be very useful to us,
\begin{equation}
 s_{k\pm}^r(i,j)=r^{\textnormal{th}} \textnormal{ digit of } \sum_{p=i}^{j-1}d_{k\pm}^p.
\label{sumstrings}
\end{equation}
This sum tells us how many turns our protein has taken in the $k\pm$ direction between any two amino acids. For instance, $s^{1}_{x-}(3,9)$ would tell us the value of the 1st binary digit of an integer representing the number of times that the protein turned in the negative $x$ direction (aka left) between amino acids 3 and 9. While the size of the output of the circuit given in Fig.~\ref{genSum} scales exactly with the size of the input, the maximum number of bits needed to represent the sum of a set of bits scales logarithmically; therefore, many of the bits representing higher places in the sequence are zero. Specifically, the sum of $n$ bits requires at most $\lceil\log_{2} n\rceil$ bits to represent in binary.

\subsubsection{Construction of $E_{overlap}\left(\boldsymbol{q}\right)$ with circuit}

The overlap penalty should be positive if any two amino acids are at the same lattice point. For a pair $i,j$, this occurs when the number of turns between them in each direction $k\pm$ is equal to those in the opposite direction $k\mp$ or equivalently, when the bit-strings representing those numbers, $s_{k+}^{j-i}\cdots s_{k+}^1$ and $s_{k-}^{j-i}\cdots s_{k-}^1$, are the same. As discussed above, since only the first $\lceil\log_{2}(j-i)\rceil$ digits of $s_{k\pm}$ are non-zero, the overlap penalty function for amino acids $i$,$j$ is
\begin{equation}
E_{overlap}(i,j)=\prod_{k=1}^D\left(\prod_{r=1}^{\lceil\log(j-i)\rceil} \textnormal{XNOR}\left(s_{k+}^r(i,j),s_{k-}^r(i,j)\right)\right),
\end{equation}
where
\begin{equation}
 \textnormal{XNOR}(p,q)=1-p-q+2pq
\end{equation}
is the exclusive {\sc nor} function which evaluates to {\sc true} if and only if the two bits have the same value. Furthermore, we need not consider every pair of amino acids in the sequence because in order for the number of turns in opposite directions to be equal, there must be an even number of total turns. The total on-site penalty function is
\begin{equation}\label{eq:OnsiteTotal}
 E_{overlap}=\lambda_{overlap}\sum_{i=1}^{N-2}\left(\sum_{j=1}^{\lfloor(N-i)/2\rfloor}E_{overlap}(i,i+2j)\right)
\end{equation}

\subsubsection{Construction of $E_{pair}(\boldsymbol{q})$ with circuit}

To determine if a pair of amino acids is adjacent on the lattice without using ancilla bits is more involved. Two amino acids are adjacent if and only if the number of turns between them in opposite directions is the same in all but one dimension and the numbers of turns in the other dimension have a difference of one. The construction of the equality condition is the same as in as for the overlap function; to construct the latter condition, consider the set of 4 bit numbers, as shown in Figure \ref{4bits}.
\begin{figure}[h]
  \begin{center}
   \begin{tabular}{|c|c|}
	\hline
	Decimal & Binary\\\hline
	0&0000\\
	1&0001\\
	2&0010\\
	3&0011\\
	4&0100\\
	5&0101\\
	6&0110\\
	7&0111\\
	8&1000\\
	9&1001\\
	10&1010\\
	11&1011\\
	12&1100\\
	13&1101\\
	14&1110\\
	15&1111\\\hline
    \end{tabular}
  \end{center}
  \caption{All 4 bit binary numbers and their decimal representations.}
  \label{4bits}
\end{figure}

Note that when the first of two sequential binary is even, the Hamming distance between those bit-strings are the same except for the least significant bit, e.g. 0000 and 0001, 1000 and 1001, 1110 and 1111. On the other hand, sequential numbers for which the lesser one is odd differ in at least two places, depending on where the rightmost 0 is in the lesser number, i.e.
\begin{equation}
 \begin{array}{cl}
    &00000000000\cdots 01\\
   +&**\cdots **011\cdots 11           \\ \hline
    &**\cdots **100\cdots 00
 \end{array},
\end{equation}
as in 0011 and 0100, 0111 and 1000, and 1011 and 1100.

Let us use $p$ to denote the position of the rightmost 0 in the odd, lesser number of this comparison. There are three portions of the bit strings which need attention when comparing adjacency in this case. First, all digits from the least significant and up to $p$ need to be different. Second, all digits after $p$ need to be the same. Third, within each possible adjacency direction ($k+$ or $k-$) there needs to be a change from $p-1$ to $p$. Finally, all the digits from the least significant up to the $(p-2)$th digit need to be the same. Using these conditions, for both cases when the lesser number is either even or odd, results in the adjacency terms for each of the two dimensions and all of the possible amino acid pairs, $a_{k}(i,j)$:
\small
\begin{align}
a_{k}(i,j) = & \left[\prod_{w\neq k}\left(\prod_{r=1}^{\lceil\log(j-i)\rceil} \textnormal{XNOR}\left(s_{w+}^r(i,j),s_{w-}^r(i,j)\right)\right)\right]\\
* & \left[\textnormal{XOR}\left(s_{k+}^1(i,j),s_{k-}^1(i,j)\right)\prod_{r=2}^{\lceil\log(j-i)\rceil}\textnormal{XNOR}\left(s_{k+}^r(i,j),s_{k-}^r(i,j)\right)\right.\nonumber\\
+ & \sum_{p=2}^{\lceil\log(j-i)\rceil}\left(\textnormal{XOR}\left(s_{k+}^{p-1}(i,j),s_{k+}^p(i,j)\right)\prod_{r=1}^{p-2}\textnormal{XNOR}\left(s_{k+}^{r}(i,j),s_{k+}^{r+1}(i,j)\right)\right.\nonumber\\
* & \left.\left.\prod_{r=1}^p\textnormal{XOR}\left(s_{k+}^r(i,j),s_{k-}^r(i,j)\right)\prod_{r=p+1}^{\lceil\log(j-i)\rceil}\textnormal{XNOR}\left(s_{k+}^r(i,j),s_{k-}^r(i,j)\right)\right)\right].\nonumber
\end{align}
\normalsize

Thus total contribution of the interaction between two amino acids to the total energy function is given by
\begin{equation}
 E_{pair}(i,j)=J_{ij}\left[a_x(i,j)+a_y(i,j)\right],
\end{equation}
where $J_{ij}$ is the adjacency matrix giving the energy of pairwise interactions that we used earlier. As was the case with the overlap penalty function, we need not consider all pairs of amino acids. In order for two amino acids to be adjacent there must be an odd number of turns between them, excluding the trivial case of amino acids that are adjacent in the primary sequence. Accordingly, the total pairwise interaction function is
\begin{equation}\label{eq:pwTotal}
 E_{pair}=\sum_{i=1}^{N-3}\left(\sum_{j=1}^{\lfloor(N-i-1)/2\rfloor}E_{pair}(i,1+i+2j)\right).
\end{equation}

\newpage
\section{The ``Diamond'' Encoding of SAWs}
There are many different ways in which one could encode a SAW into binary. Of all the alternatives to the ``turn'' encoding that we have considered, one stands out for a number of reasons: the so-called ``diamond encoding'' lends itself to an energy function which is natively 2-local (without any reductions) and which has a very sparse QUBO (quadratic unconstrained binary optimization) matrix. Despite the fact that the diamond encoding requires no ancillary bits whatsoever, the encoding is still less bit-wise efficient than the ``turn encoding''. In the language of constraint satisfaction programming, this means that the clause:variable ratio is significantly lower when compared to the clause:variable in the turn encoding.

\subsection{Embedding physical structure}
\begin{figure}[h]
  \centering
    \includegraphics[scale=0.3]{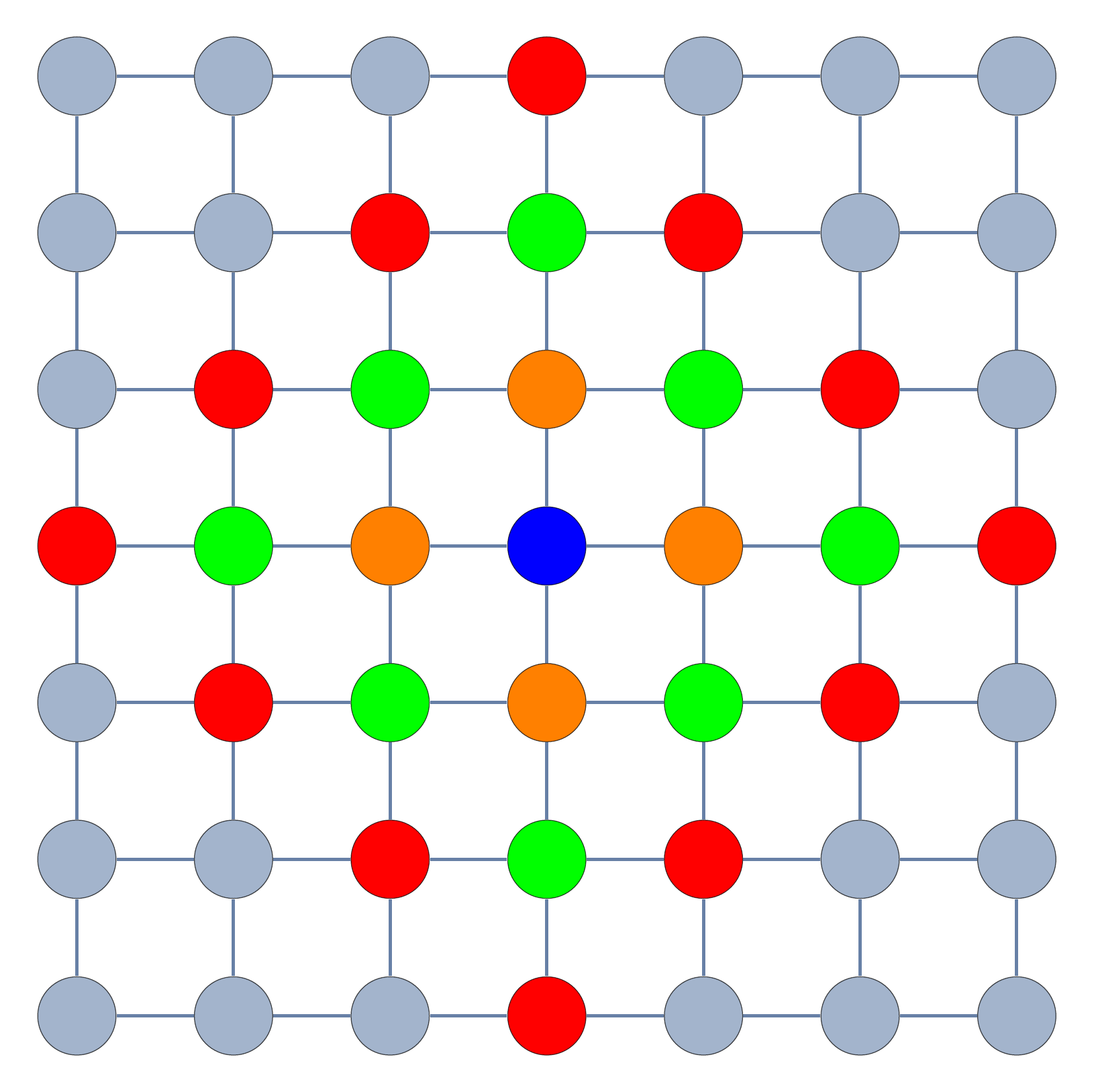}
  \caption{A ``map'' of the diamond encoding in 2D. If the first amino acid is fixed to the blue lattice point in the center then the second amino acid must be on an orange lattice point, the third must be on a green lattice point and the fourth must be on either an orange or red lattice point.}
\label{diamond}
\end{figure}

The diamond encoding can be thought of as a more sophisticated and subtle version of the ``spatial'' encoding used in \cite{Perdomo2008}. The key insight behind the diamond encoding is that if the first amino acid is fixed then each subsequent amino can only occupy a very restricted set of lattice points which can be enumerated independent of any knowledge of the particular fold. To clarify this point and elucidate why we refer to this as the ``diamond'' encoding, see Fig.~\ref{diamond}.

Fig.~\ref{diamond} illustrates what the ``diamond'' of valid lattice sites looks like for the first 4 amino acids in a SAW. In the diamond encoding each bit refers to a specific lattice site which could be occupied by an amino acid in that part of the sequence. In Fig.~\ref{diamond} we notice that the second amino acid may occupy 4 positions, the third may occupy 8 and the fourth may occupy 16. Accordingly, we need this many bits for each amino acid.

\begin{equation}
\boldsymbol{q} = \underbrace{q_1 q_2 q_3 q_4}_{2\textrm{nd acid}} \underbrace{q_5 q_6 q_7 q_8 q_9 q_{10} q_{11} q_{12}}_{3\textrm{rd acid}} \cdots
\end{equation}

Though very straightforward to encode, this representation makes significantly less efficient use of bits than does the turn representation. However, there are a few tricks which we can use to improve the situation for this encoding. While the ``diamond'' of possible lattice locations for each amino acid grows quadratically with the length of the chain we can simultaneously save bits and drastically reduce the solution space without discarding the global minimum by deciding to set a hard limit on the size of the diamond. For instance, if a protein has length 100 then we would expect that the diamond for the 100th amino acid will have a radius of 99 lattice points at each corner. However, we can use the observation that proteins always fold into very compact structures to justify a substantial restriction on the solution space of our problem.

The very fact that these problems are typically called ``protein folding'' suggests that low energy solutions involve dense conformations. Indeed, almost all heuristic methods for folding proteins take advantage of the compact nature of low energy folds to constrain search procedures \cite{Baker2000,Oakley2011,Shakhnovich1996}. A large part of the reason why lattice heteropolymer problems such as protein folding are so difficult and poorly suited to heuristic algorithms is because the low energy solutions are always very compact and thus, frustrated, which makes it very unlikely that compact folds will be found efficiently via stochastic searches \cite{Dill1993,Shea2000,Camacho1995,Nymeyer1998}. Therefore, for any interesting problem its reasonable to assume that the protein will not stretch out further than a certain limit. To estimate this limit one must have familiarity with the types of solutions expected of the particular problem. An examination of several publications holding current records for lowest energy folds in canonical problems suggests to us that for a 100 unit instance in $2D$ a reasonable cutoff radius would be around 20-30 lattice points. The cutoff radius could reasonably be made shorter for lattice models in higher dimensions as folds are expected to be even more compact on higher dimensional lattices. The number of bits required for the diamond encoding can be expected to grow cubicly up to a limit and then linearly after that limit if a cutoff is imposed. Because the number of bits required for the turn ancilla grows quadratically, for large proteins or proteins on higher dimensional lattices the diamond encoding would actually be more efficient in bit resources.

\subsection{Natively 2-local $E(\boldsymbol{q})$}

The major advantages of the diamond encoding become evident as soon as one starts to construct $E(\boldsymbol{q})$. The breakdown of the energy function looks different for the diamond encoding than for the turn encoding because the diamond encoding has different strengths and weaknesses. The first difference is that the diamond encoding will require a constraint, $E_{one}\left(\boldsymbol{q}\right)$ which makes sure that each amino acid will have only one bit flipped to ``on'' so that each amino acid can only occupy one lattice position. Furthermore, the diamond encoding does not hard-code a primary structure constraint so we will need a term, $E_{connect}\left(\boldsymbol{q}\right)$ to guarantee that each sequential amino acid is adjacent. Like the turn encoding the diamond encoding will also require $E_{overlap}\left(\boldsymbol{q}\right)$ and $E_{pair}\left(\boldsymbol{q}\right)$ terms. The overall energy function looks like,
\begin{equation}
E\left(\boldsymbol{q}\right) = E_{one}\left(\boldsymbol{q}\right) + E_{connect}\left(\boldsymbol{q}\right) + E_{overlap}\left(\boldsymbol{q}\right) + E_{pair}\left(\boldsymbol{q}\right).
\end{equation}

\subsubsection{Construction of $E_{one}\left(\boldsymbol{q}\right)$}
Each amino acid is encoded by flipping a bit in the part of the total bit-string sequence which represents that amino acid. Thus, we need to make sure that exactly one bit is flipped ``on'' for each amino acid. The most efficient way to guarantee this is the case for low energy solutions is to lower the energy whenever a bit is flipped on but introduce extremely high penalties if any two are flipped on for the same amino acid. For instance, if $\boldsymbol{q}^{k}$ is the binary vector which represents the $k$th amino acid and $n_{k}$ represents the length of this vector then we can write,
\begin{equation}
E_{one}\left(\boldsymbol{q}\right) = \lambda_{one} \sum^{N}_{k=2} \sum^{n_{k}-1}_{i=1}\sum^{n_{k}}_{j>i} q_{i}^{k} q_{j}^{k}.
\label{one}
\end{equation}
$\lambda_{one}$ in Eq.~\ref{one} yields terms which impose a very large penalty if any two (or more) bits are flipped at once. As written, this function allows for the possibility that no bits are flipped on at once (and clearly one must be flipped on). However, the terms introduced in $E_{connect}\left(\boldsymbol{q}\right)$ will guarantee that the low energy solutions all have one bit flipped on. Thus, this function only needs to make sure that no more than one bit is flipped for each amino acid.

\subsubsection{Construction of $E_{connect}\left(\boldsymbol{q}\right)$}
To form $E_{connect}\left(\boldsymbol{q}\right)$ we take a very similar approach to how we formed $E_{one}\left(\boldsymbol{q}\right)$. To guarantee that the low energy solution space contains only amino acids chains which connect in the desired order we couple every bit representing amino acid $k$ to each of the (at most 4) bits representing a lattice position adjacent to that amino acid from the previous amino acid $k-1$ and multiply by a reward as follows (using the same notation as was used in Eq.~\ref{one},
\begin{equation}
E_{connect}\left(\boldsymbol{q}\right) = \lambda_{connect}(N-2)-\lambda_{connect} \sum^{N}_{k=3} \sum_{i=1}^{n_k} \sum_{j=1}^{n_{k-1}} q^{k}_{i} q^{k-1}_{j}a(i,k,j,k-1),
\label{con}
\end{equation}
where $a(i,k,j,l)$ is an adjacency indicator, i.e. is one if the lattice points corresponding to $q_i^k$ and $q_j^l$ are adjacent and is zero otherwise.
One important caveat is that $\lambda_{connect} << \lambda_{one}$ so that the system cannot overcome the $\lambda_{one}$ penalty by having multiple $\lambda_{connect}$ couplings. Finally we put the constant factor of $N-2$ into the equation to adjust the energy back to zero overall for valid solutions which contain $N-2$ connections.

\subsubsection{Construction of $E_{overlap}\left(\boldsymbol{q}\right)$}
It is much easier to prevent amino acids from overlapping in the diamond mapping than in the turn mapping. The only way that amino acids could overlap in the diamond mapping is for amino acids which have an even number of bonds between them to flip bits corresponding to the same lattice location. For instance, in Fig.~\ref{diamond} its clear that the fourth amino acid could overlap with second amino acid since the orange lattice points are possibilities for both amino acids. Assuming that the diamond lattice positions are encoded so that the inner diamond bits come first in the bit-string for each amino acid and that bits are enumerated in some consistent fashion (e.g. starting at the top and going clockwise around the diamond), we can write the following,
\begin{equation}
E_{overlap}\left(\boldsymbol{q}\right) = \lambda_{overlap} \sum^{N-1}_{k=2} \sum^{N}_{h>k} \sum_{i=1}^{n_{k}} \left[\left(1+k-h\right) \textrm{mod}\, 2\right] q^{k}_{i} q^{h}_{i}.
\end{equation}
This expression would perfectly sum over all the possible overlaps as the first two sums iterate through all possible overlapping pairs and the third sum iterates through all of the diamond points up to the last point they both share, $n_{k}$.

\subsubsection{Construction of $E_{pair}\left(\boldsymbol{q}\right)$}
To form the pairwise interaction term we simply couple each bit to the possible adjacent lattice locations which could be occupied by other amino acids. The strength of the coupling will depend on the interaction matrix element between the two amino acids coupled by the term. Additionally, we note that amino acids are only able to be adjacent if there are an even number of amino acids (2 or greater) in between the two. Thus, the formula is as follows:
\begin{equation}
E_{pair}\left(\boldsymbol{q}\right) = \sum^{N-3}_{k=2} \sum^{N}_{h=k+3} \sum_{<i j>} J_{hk} \left[\left(k-h\right) \textrm{mod}\, 2\right] q^{k}_{i} q^{h}_{j}
\end{equation}
where the sum over $<i j>$ is understood as a sum over bits corresponding to adjacent lattice sites. There is no straightforward way to write the function $<i j>$ in analytical terms. Nevertheless, for large problems it is trivial to write a program which iterates through bits in the second amino acid with a for-loop and evaluated the sum on those bits if the first amino acid bit and the second amino acid bit have a grid distance of 1.

\newpage
\section{Pseudo-boolean Function to W-SAT}

In order to take advantage of state-of-the-art \textsc{satisfiability} (SAT) solvers to optimize our pseudo-boolean function, it is necessary to map the problem to \textsc{Weighted Maximum Satisfiability} (W-SAT). The most general form of the generic SAT problem is known as K-SAT. In K-SAT the problem is to find a vector of boolean valued variables which satisfies a list of clauses, each containing up to K variables, which constrain the solution. When K-SAT has a solution it is known as ``satisfiable'' and for K $\leq$ 2 the problem is tractable in polynomial time. However, for K $>$ 2 the problem is known to be \textsc{NP-complete}; in fact, 3-SAT was the first problem proved to be \textsc{NP-Complete} \cite{Cook1971a}.

\subsection{MAX-SAT and W-SAT}
\label{SATexample}

\textsc{Maximum Satisfiability} (MAX-SAT) is a more difficult version of the canonical SAT problem which is relevant when K-SAT is either ``unsatisfiable'' or at least not known to be satisfiable. In MAX-SAT the goal is not necessarily to find the solution string which satisfies all clauses (such a solution string may not even exist); rather, the goal is to find the solution string which satisfies the maximum number of clauses.

An extension of MAX-SAT known as \textsc{Weighted Maximum Satisfiability} (aka W-SAT) is what will be most relevant to us. In W-SAT each clause is given a positive integer valued ``weight'' which is added to a sum only if the clause evaluates to {\sc false}. Accordingly, in W-SAT the goal is to minimize this sum rather than the total number of {\sc false} clauses as in canonical MAX-SAT \cite{Xing2005,Boros2002}. We can more succinctly state the problem as follows: given $m$ number of clauses ($y$) each with a weight of $w$, minimize
\begin{equation}
W = \sum_{i=1}^{m} w_{i} y_{i} \,\,\, : \,\,\, y_{i} = \begin{cases}
1 & \textrm{if the $i$th clause is {\sc false}} \\
0 & \textrm{otherwise}.
\end{cases}
\end{equation}

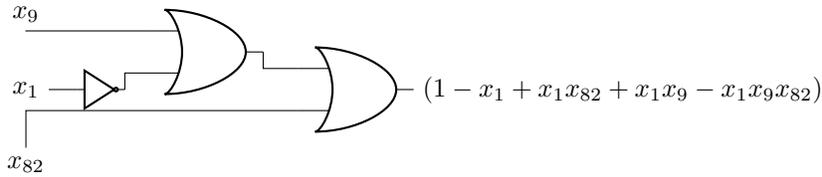
\begin{figure}[h]
\centering
\begin{circuitikz}
\draw
node (x9) at (0,2) {$x_{9}$}
node (x1) at (0,1) {$x_{1}$}
node (x82) at (0,0) {$x_{82}$}

node[scale=0.45, not port] (not1) at (1,1) {}
node[or port] (or1) at (3,1.5) {}
node[or port] (or2) at (5,1) {}

(x1) -- (not1.in)
(not1.out) |- (or1.in 2)
(x9) |- (or1.in 1)
(or1.out) |- (or2.in 1)
(x82) |- (or2.in 2)
(or2.out) node[right] {{($1-x_{1}+ x_{1} x_{82}+x_{1} x_{9} - x_{1}x_{9} x_{82} $})};
\end{circuitikz}
\caption{A logical circuit representation of the CNF clause: ``$9\,\,\,-\!1\,\,\,82$''}
\label{cnfcir}
\end{figure}

The same approximation schemes and exact solver algorithms which work for MAX-SAT also work for W-SAT \cite{Boughaci2004,Pankratov2010}. In order to use these solvers one must first translate their pseudo-boolean function into a W-SAT problem articulated in what is known as Weighted Conjunctive Normal Form (WCNF). In WCNF, the W-SAT problem is stated as a list of weights followed by a clause with each clause stated as an {\sc or} statement between integers representing the index of the corresponding boolean variable in the solution vector. In WCNF, a negative integer denotes a negation. For instance the WCNF clause ``$4000\,\,\,9\,\,\,-\!1\,\,\,82$'' means $x_{9} \vee \neg x_{1} \vee x_{82}$ with penalty of $4000$ if clause evaluates to {\sc false}. Fig.~\ref{cnfcir} shows this clause as a logic circuit.

\subsection{Constructing WCNF clauses}
To prepare the WCNF input file from a pseudo-boolean function one will need to write a short script which transforms each term in the pseudo-boolean function into a WCNF clause. There is more than one way to accomplish this transformation and we will only discuss one method here. For a more complete review of this topic, see \cite{Een2006}.

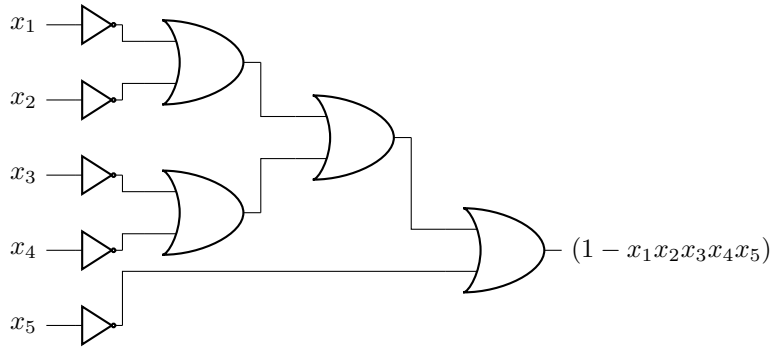
\begin{figure}[h]
\centering
\begin{circuitikz}
\draw
node (x1) at (0,4) {$x_{1}$}
node (x2) at (0,3) {$x_{2}$}
node (x3) at (0,2) {$x_{3}$}
node (x4) at (0,1) {$x_{4}$}
node (x5) at (0,0) {$x_{5}$}

node[scale=0.45, not port] (not1) at (1,4) {}
node[scale=0.45, not port] (not2) at (1,3) {}
node[scale=0.45, not port] (not3) at (1,2) {}
node[scale=0.45, not port] (not4) at (1,1) {}
node[scale=0.45, not port] (not5) at (1,0) {}
node[or port] (or1) at (3,1.5) {}
node[or port] (or2) at (3,3.5) {}
node[or port] (or3) at (5,2.5) {}
node[or port] (or4) at (7,1) {}

(x1) -- (not1.in)
(x2) -- (not2.in)
(x3) -- (not3.in)
(x4) -- (not4.in)
(x5) -- (not5.in)
(not1.out) |- (or2.in 1)
(not2.out) |- (or2.in 2)
(not3.out) |- (or1.in 1)
(not4.out) |- (or1.in 2)
(not5.out) |- (or4.in 2)
(or1.out) |- (or3.in 2)
(or2.out) |- (or3.in 1)
(or3.out) |- (or4.in 1)
(or4.out) node[right] {{($1-x_{1} x_{2} x_{3} x_{4} x_{5} $})};
\end{circuitikz}
\caption{A logical circuit which shows that any pseudo-boolean term with positive weight is equivalent (up to a constant) to a CNF clause with each variable negated. The term produced here is negative because the weight is only added when the clause evaluates to {\sc false}.}
\label{cir2}
\end{figure}

It will be very useful to think of CNF clauses as logical circuits which involve only {\sc or} gates and {\sc not} gates as in Fig.~\ref{cnfcir}. Weights in WCNF notation always represent a positive value. Because pseudo-boolean functions are treated as cost functions to minimize and the goal of W-SAT is to minimize the sum of weights on {\sc false} clauses, terms in the pseudo-boolean function with a positive weight are very easy to translate in WCNF notation. To achieve this, one needs only to pass all variables in the clause through a {\sc not} gate and then a series of {\sc or} gates (effectively making a {\sc nand} gate which takes all variables as input). This circuit is illustrated in Fig.~\ref{cir2} for the case of a 5 variables clause.

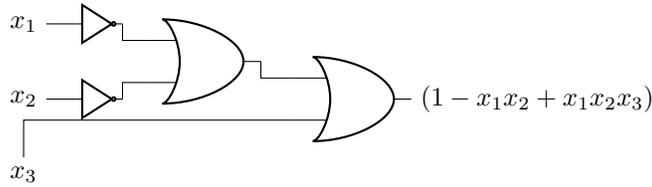
\begin{figure}[h]
\centering
\begin{circuitikz}
\draw
node (x1) at (0,2) {$x_{1}$}
node (x2) at (0,1) {$x_{2}$}
node (x3) at (0,0) {$x_{3}$}

node[scale=0.45, not port] (not1) at (1,2) {}
node[scale=0.45, not port] (not2) at (1,1) {}
node[or port] (or1) at (3,1.5) {}
node[or port] (or2) at (5,1) {}

(x1) -- (not1.in)
(x2) -- (not2.in)
(not1.out) |- (or1.in 1)
(not2.out) |- (or1.in 2)
(or1.out) |- (or2.in 1)
(x3) |- (or2.in 2)
(or2.out) node[right] {{($1-x_{1} x_{2}+x_{1} x_{2} x_{3} $})};
\end{circuitikz}
\caption{A logical circuit on three variables which gives a positive valued 3-local CNF term.}
\label{cir3}
\end{figure}

\begin{figure}[h]
\centering
\begin{circuitikz}
\draw
node (x1) at (0,1) {$x_{1}$}
node (x2) at (0,0) {$x_{2}$}

node[scale=0.45, not port] (not1) at (1,1) {}
node[or port] (or1) at (3,0.5) {}

(x1) -- (not1.in)
(not1.out) |- (or1.in 1)
(x2) |- (or1.in 2)
(or1.out) node[right] {{($1-x_{1} + x_{1} x_{2}$})};
\end{circuitikz}
\caption{A logical circuit on three variables which gives a positive valued 2-local CNF term.}
\label{cir4}
\end{figure}
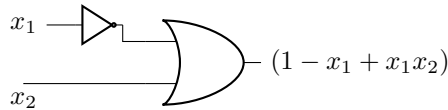

Representing a negative weighted pseudo-boolean term in CNF is less trivial but follows a simple pattern. To make the CNF clause positive (corresponding to negative boolean term) one needs to construct the same circuit as in the case when the boolean term is positive but remove one of the {\sc not} gates. An example comprising three variables is shown in Fig.~\ref{cir3}.
However, this circuit alone does not accomplish our goal as it produces a 2-local term with negative weight in addition to the 3-local term with positive weight. Consequentially, after using the circuit in Fig.~\ref{cir3} to get rid of the 3-local term ``$x_{1} x_{2} x_{3}$'' we must subtract the term ``$x_{1} x_{2}$'' multiplied by its weight from the pseudo-boolean expression we are converting into CNF. At first glance, it is not obvious that this procedure will get us anywhere - we turned a term into CNF only to introduce a new term into the pseudo-boolean which we must convert back into CNF. However, the auxiliary terms produced by this circuit are of one degree less than number of variables in the term; thus, we can iterate this procedure until only the constant term remains. The next CNF clause (this time 2-local) is shown in Fig.~\ref{cir4}.

\subsection{Solving SAT problems}
While MAX-SAT is known to be \textsc{NP-hard}, there exist heuristic algorithms which are guaranteed to satisfy a fixed fraction of the clauses of the optimal solution in polynomial time. In general, oblivious local search will achieve at least an approximation ratio of $\frac{k}{k+1}$, Tabu search achieves a ratio of at least $\frac{k+1}{k+2}$ and non-oblivious local search achieves an approximation ratio of $\frac{2^k-1}{2^k}$ where $k$ is the ``K'' in K-SAT. For the special case of MAX-2-SAT the best possible algorithm is theoretically capable of satisfying at least $\frac{21}{22} + \epsilon \approx 0.955 + \epsilon \left[9\right]$ in polynomial time \cite{Pankratov2010,Choi2009}. Additionally, there are a great deal of exact MAX-SAT solvers which run in super-polynomial time but in many cases can find the solution to MAX-SAT in a very short amount of time, even for problems containing hundreds of variables and clauses \cite{Marques-Silva2007,Larrosa2006}.

\newpage
\section{W-SAT to Integer-Linear Programming}

Integer-Linear Programming (ILP) is a subset of linear programming problems in which some variables are restricted to integer domains. In general, ILP is an NP-Hard problem but the importance of ILP problems (particular for logistics scheduling) has produced many extremely good exponential-time exact solvers and polynomial-time heuristic solvers \cite{Xing2005}. Pseudo-boolean optimization is an even more specific case of ILP sometimes known as 0-1 ILP where the integer variables are boolean \cite{Boros2002}. The mapping between W-SAT and ILP is very straightforward.

\subsection{Mapping to ILP}

In ILP, the goal is to minimize an objective function of integer-valued variables subject to a list of inequality constraints which must be satisfied. The inequality constraints come directly from the clauses in W-SAT. As described in Sec.~\ref{SATexample}, the logical clause from the WCNF clause ``4000 9 -1 82'' (which again, means $x_{9} \vee \neg x_{1} \vee x_{82}$ with penalty of $4000$ if clause evaluates to False) can be represented as $x_9 + (1-x_1) + x_{82} \geq 1$ s.t. $x_n \in \{0, 1\}$. In ILP, all constraints must be satisfied but in W-SAT clauses are sometimes not satisfied; to accommodate this we introduce an auxiliary binary variable, $y_1$ into the equation and get $y_1 + x_9 + (1-x_1) + x_{82} \geq 1$. Thus, if the original equation is False, $y_1$ will have a value of True which satisfies the inequality. We can take advantage of this auxiliary variable to construct the optimization function, $W$. Since the clause in our example has a weight of $4000$ we can write $W = 4000 y_1$ s.t. $y_1 + x_9 + (1-x_1) + x_{82} \geq 1$. Thus, the mapping between ILP and W-SAT is extremely trivial: all WCNF clauses are rewritten as linear equalities which are $\geq 1 - y_i$ by adding together the variables (or their negations) where $i$ is the index of the clause and the objective function is written as $W = \sum_{i=1}^{N} w_{i} y_{i}$ where $N$ is the number of clauses and $w_{i}$ is the weight of that clause \cite{Xing2005}.

\subsection{Solving ILP problems}
Commercial logistic scheduling software such as IBM ILOG CPLEX Optimization Studio (aka CPLEX) is designed to solve in integer programming, linear programming, and mixed integer-linear programming problems on a very large scale \cite{cplex}. Constraint satisfaction problems which are sometimes very difficult to solve using conventional SAT techniques can be easier to solve using ILP techniques and vice versa. In particular, SAT solvers and specialized pseudo-boolean optimizers seem to outperform ILP solvers when a problem is over-constrained \cite{Aloul2002}. On the other hand, for problems which are under-constrained and have a large number of variables ILP solvers are the natural choice. In some cases 0-1 ILP optimizers such as Pueblo will outperform both SAT solvers and commercial ILP solvers \cite{Sakallah2006,Sheini2005,Manolios2011}.

\newpage
\section{Locality Reductions}
\label{reduction}

The practical ability to either exactly or approximately solve random instances of constraint satisfaction optimization such as pseudo-boolean optimization or MAX-SAT seems to depend very sensitively on the variable to clause ratio and degree of constraint expressions \cite{Pankratov2010,Kahl2011,Xing2005}. In fact, the degree of constraints determines the complexity class of certain constraint satisfaction problems; e.g. 2-SAT is proven to be in P whereas 3-SAT is in \textsc{NP-Complete} \cite{Cook1971a}. Clearly for instances such as this there can be no efficient method which reduces the degree of constraints. Fortunately, reducing the degree of constraints in general pseudo-boolean optimization (i.e. reducing the polynomial order of pseudo-boolean terms) can be done efficiently.

Constraint degree reduction is particularly important if we wish to solve our problem using existing architectures for adiabatic quantum computation because available devices tend to be very limited in their ability to realize arbitrary variable couplings (especially high ordered couplings). For instance, the D-Wave One device used for pseudo-boolean optimization in \cite{Perdomo-Ortiz2012} is only able to implement 2-local qubit couplings and has limited coupler resolution. To encode functions of higher locality in such setups, we must introduce ancilla bits which replace 2-local terms to reduce locality. Because these ancilla become free parameters of the system, it is also necessary to introduce penalty functions to account for the possibility that their value may be incorrect. All of this is accomplished with the function $E_\wedge \left(q_i, q_j, \tilde q_n;\delta_n\right)$ in Eq.~\ref{reduce} which introduces the ancillary bit $q_n$ in order to collapse the 2-local term $q_i q_j$ with energy penalty of $\delta_n$ if $q_n \neq q_i q_j$. For a further discussion, see \cite{Biamonte2008,Perdomo2008,Babbush2013b}.

\begin{equation}
E_\wedge (q_i, q_j, \tilde q_n;\delta_n) = \delta_n (3 \tilde q_n + q_i q_j - 2 q_i \tilde q_n - 2 q_j \tilde q_n)
\label{reduce}
\end{equation}

If one desires an entirely 2-local energy function then many $E_\wedge \left(q_i, q_j, \tilde q_n;\delta_n\right)$'s may be necessary to collapse all high-local terms. For instance, consider the complete energy function for the HP model protein $H P P H P$ when coded in the turn ancilla mapping:
\small
\begin{align}
E & = -4 q_2 q_6 \lambda _1+4 q_1 q_3 q_6 \lambda _1+3 q_6 \lambda _1+28 q_1 \lambda _2+25 q_1 q_2 \lambda _2+108 q_2 \lambda _2-56 q_1 q_3\lambda _2 \label{hpphp}\\
& -50 q_1 q_2 q_3 \lambda _2+26 q_2 q_3 \lambda _2+28 q_3 \lambda _2+24 q_1 q_4 \lambda _2-16 q_1 q_2 q_4 \lambda _2-56 q_2 q_4 \lambda _2-48 q_1 q_3 q_4 \lambda _2 \nonumber\\
& +32 q_1 q_2 q_3 q_4 \lambda _2-18 q_2 q_3 q_4 \lambda _2+25 q_3 q_4 \lambda _2+108 q_4 \lambda _2-56 q_1 q_5 \lambda_2-48 q_1 q_2 q_5 \lambda_2 \nonumber\\
& +25 q_2 q_5 \lambda _2+48 q_1 q_3 q_5 \lambda _2-50 q_2 q_3 q_5 \lambda _2-56 q_3 q_5 \lambda _2-48 q_1 q_4 q_5 \lambda _2+32 q_1 q_2 q_4 q_5 \lambda _2 \nonumber\\
& -18 q_2 q_4 q_5 \lambda _2+36 q_2 q_3 q_4 q_5 \lambda _2-50 q_3 q_4 q_5 \lambda _2+25 q_4 q_5 \lambda _2+28 q_5 \lambda _2-32 q_1 q_7 \lambda _2 \nonumber\\
& -96 q_2 q_7 \lambda _2+64 q_1 q_3 q_7 \lambda _2-32 q_3 q_7 \lambda _2+64 q_2 q_4 q_7 \lambda_2-96 q_4 q_7 \lambda _2+64 q_1 q_5 q_7 \lambda _2 \nonumber\\
& +64 q_3 q_5 q_7 \lambda _2-32 q_5 q_7 \lambda _2-32 q_7 \lambda _2-16 q_1 q_8 \lambda _2-48 q_2 q_8 \lambda _2+32 q_1 q_3 q_8 \lambda _2-16 q_3 q_8 \lambda _2 \nonumber\\
& +32 q_2 q_4 q_8 \lambda _2-48 q_4 q_8 \lambda _2+32 q_1 q_5 q_8 \lambda _2+32 q_3 q_5 q_8 \lambda _2-16 q_5 q_8 \lambda _2+64 q_7 q_8 \lambda _2 \nonumber\\
& -32 q_8 \lambda _2-8 q_1 q_9 \lambda _2-24 q_2 q_9 \lambda _2+16 q_1 q_3 q_9 \lambda _2-8 q_3 q_9 \lambda _2+16 q_2 q_4 q_9 \lambda _2-24 q_4 q_9 \lambda _2 \nonumber\\
& +16 q_1 q_5 q_9 \lambda _2+16 q_3 q_5 q_9 \lambda _2-8 q_5 q_9 \lambda _2+32 q_7 q_9 \lambda _2+16 q_8 q_9 \lambda _2-20 q_9\lambda _2-4 q_1 q_{10} \lambda _2 \nonumber\\
& -12 q_2 q_{10} \lambda _2+8 q_1 q_3 q_{10} \lambda _2-4 q_3 q_{10} \lambda _2+8 q_2 q_4 q_{10} \lambda _2-12 q_4 q_{10} \lambda _2+8 q_1 q_5 q_{10} \lambda _2 \nonumber\\
& +8 q_3 q_5 q_{10}\lambda _2-4 q_5 q_{10} \lambda _2+16 q_7 q_{10} \lambda _2+8 q_8 q_{10} \lambda _2+4 q_9 q_{10} \lambda _2-11 q_{10} \lambda _2+36 \lambda _2.\nonumber
\end{align}
\normalsize

In order to reduce this function to 2-local we will need to collapse some of the 2-local terms inside of the 3-local terms to a single bit. We enumerate all of the 3-local terms and their corresponding 2-local terms which we could use to reduce each 3-local term in Eq.~\ref{combos}.
\tiny
\begin{equation}
\label{combos}
\begin{aligned}[c]
\left(
\begin{array}{ccc}
 q_1 & q_2 & q_3 \\
 q_1 & q_2 & q_4 \\
 q_1 & q_3 & q_4 \\
 q_2 & q_3 & q_4 \\
 q_1 & q_2 & q_3 \\
 q_1 & q_2 & q_5 \\
 q_1 & q_3 & q_5 \\
 q_2 & q_3 & q_5 \\
 q_1 & q_4 & q_5 \\
 q_2 & q_4 & q_5 \\
 q_1 & q_2 & q_4 \\
 q_3 & q_4 & q_5 \\
 q_2 & q_3 & q_4 \\
 q_1 & q_3 & q_6 \\
 q_1 & q_3 & q_7 \\
 q_2 & q_4 & q_7 \\
 q_1 & q_5 & q_7 \\
 q_3 & q_5 & q_7 \\
 q_1 & q_3 & q_8 \\
 q_2 & q_4 & q_8 \\
 q_1 & q_5 & q_8 \\
 q_3 & q_5 & q_8 \\
 q_1 & q_3 & q_9 \\
 q_2 & q_4 & q_9 \\
 q_1 & q_5 & q_9 \\
 q_3 & q_5 & q_9 \\
 q_1 & q_3 & q_{10} \\
 q_2 & q_4 & q_{10} \\
 q_1 & q_5 & q_{10} \\
 q_3 & q_5 & q_{10} \\
\end{array}
\right)
\end{aligned}
\qquad\Longleftrightarrow\qquad
\begin{aligned}[c]
\left(
\begin{array}{ccc}
 q_1 q_2 & q_1 q_3 & q_2 q_3 \\
 q_1 q_2 & q_1 q_4 & q_2 q_4 \\
 q_1 q_3 & q_1 q_4 & q_3 q_4 \\
 q_2 q_3 & q_2 q_4 & q_3 q_4 \\
 q_1 q_2 & q_1 q_3 & q_2 q_3 \\
 q_1 q_2 & q_1 q_5 & q_2 q_5 \\
 q_1 q_3 & q_1 q_5 & q_3 q_5 \\
 q_2 q_3 & q_2 q_5 & q_3 q_5 \\
 q_1 q_4 & q_1 q_5 & q_4 q_5 \\
 q_2 q_4 & q_2 q_5 & q_4 q_5 \\
 q_1 q_2 & q_1 q_4 & q_2 q_4 \\
 q_3 q_4 & q_3 q_5 & q_4 q_5 \\
 q_2 q_3 & q_2 q_4 & q_3 q_4 \\
 q_1 q_3 & q_1 q_6 & q_3 q_6 \\
 q_1 q_3 & q_1 q_7 & q_3 q_7 \\
 q_2 q_4 & q_2 q_7 & q_4 q_7 \\
 q_1 q_5 & q_1 q_7 & q_5 q_7 \\
 q_3 q_5 & q_3 q_7 & q_5 q_7 \\
 q_1 q_3 & q_1 q_8 & q_3 q_8 \\
 q_2 q_4 & q_2 q_8 & q_4 q_8 \\
 q_1 q_5 & q_1 q_8 & q_5 q_8 \\
 q_3 q_5 & q_3 q_8 & q_5 q_8 \\
 q_1 q_3 & q_1 q_9 & q_3 q_9 \\
 q_2 q_4 & q_2 q_9 & q_4 q_9 \\
 q_1 q_5 & q_1 q_9 & q_5 q_9 \\
 q_3 q_5 & q_3 q_9 & q_5 q_9 \\
 q_1 q_3 & q_1 q_{10} & q_3 q_{10} \\
 q_2 q_4 & q_2 q_{10} & q_4 q_{10} \\
 q_1 q_5 & q_1 q_{10} & q_5 q_{10} \\
 q_3 q_5 & q_3 q_{10} & q_5 q_{10} \\
\end{array}
\right)
\end{aligned}
\end{equation}
\normalsize

Eq.~\ref{combos} shows that there are 30, 3-local terms in Eq.~\ref{hpphp} and three different ways to collapse each of those 3-local terms. In general, the problem of choosing the most efficient 2-local terms to collapse this function is \textsc{NP-HARD} (see \cite{Boros2002}). This becomes evident if we represent our problem as an element cover on a bipartite graph. Suppose we relabel each 3-local term on the left as ``set'' 1-30, denoted as $S_1 S_2 ... S_{30}$. We can then make the following bipartite graph which connects the 3-local terms to the 2-local terms which collapse them.
\begin{figure}[h]
 \centering
  \includegraphics[scale=0.5]{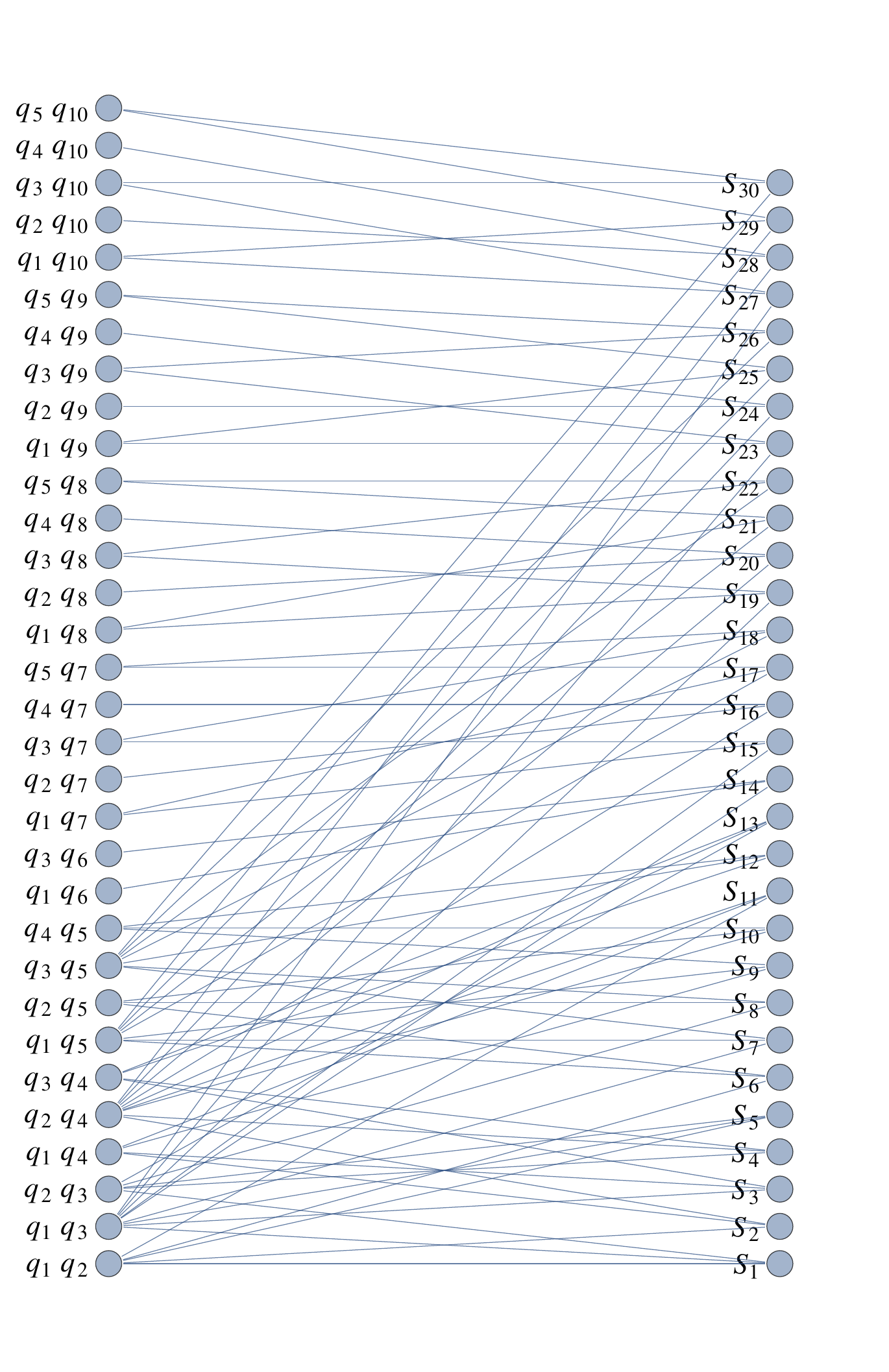}
 \caption{A bipartite graph connecting the 3-local terms ($S_{n}$) in Eq.~\ref{hpphp} to the 2-local terms ($q_{i}q_{j}$) which collapse them.}
 \label{graph}
\end{figure}

Fig.~\ref{graph} shows that we can now restate the problem in the following way: ``choose the fewest number of 2-local terms (on the left) which covers all 3-local terms (on the right) with at least one edge.'' In general, this problem is isomorphic to the canonical ``hitting set'' problem which is equivalent to set cover, one of Karp's 21 \textsc{NP-Complete} Problems \cite{Chandrasekaran2011,Chvatal1979,Laue2008}. However, we have specifically kept this issue in mind when creating the turn-ancilla representation in such a way as to guarantee that it is easy to find a relatively efficient solution to this problem. Accordingly, our experience has been that a greedy local-search algorithm performs very well \cite{Babbush2013b}.

The explanation for this is simple: each 3 or 4-local term will contain no more than 1 ancillary bit; thus, to cover all 3 and 4-local terms we can focus entirely on the physical bits (in this case, bits 1-5). In alternative mappings not presented here we have frequently encountered extremely difficult instances of the hitting set problem during the reduction process. In these situations one should see \cite{Shi2010} for a very efficient algorithm which can exactly solve hitting cover in $O\left(1.23801^{n}\right)$.

\newpage
\section{Quantum Realization}
\label{quantum}

A primary goal of this review is to elucidate an efficient process for encoding chemical physics problems into a form suitable for quantum computation. In addition to providing the alternatives for the solution of the lattice heteropolymer problem in quantum devices, we seek to provide a general explanation of considerations for constructing energy functions for these devices. These have many possible applications for solving problems related to statistical mechanics on the device. In this section, we will complete our review by demonstrating the final steps required to embed a small instance of a particular lattice protein problem into a QUBO Hamiltonian.

The Hamiltonians and the number of resources presented here correspond to the minimum amount of resources needed assuming the device can handle many-body interactions as is the case for NMR quantum computers or trapped ions. The hierarchical experimental proposals presented here work for lattice folding under no external constraints, i.e., amino acid chains in ``free space'' \footnote{External interactions could also be included as presented and verified experimentally in \cite{Perdomo-Ortiz2012}.}. As a final step we will reduce these Hamiltonians to a 2-local form specifically design for the D-Wave One used in \cite{Perdomo-Ortiz2012,Neven2008a,Denchev2012,Johnson2011}. The final Hamiltonian we present is more efficient than that used in \cite{Perdomo-Ortiz2012} as we have since realized several tricks to make the energy function more compact.

\subsection{Previous experimental implementation}
Throughout this review we have referred to an experimental implementation of quantum annealing to solve lattice heteropolymer problems in \cite{Perdomo-Ortiz2012}. The quantum hardware employed consists of 16 units of a recently characterized eight qubit unit cell \cite{Johnson2011,Harris2010}. Post-fabrication characterization determined that only 115 qubits out of the 128 qubit array can be reliably used for computation. The array of coupled superconducting flux qubits is, effectively, an artificial Ising spin system with programmable spin-spin couplings and transverse magnetic fields. It is designed to solve instances of the following (NP-hard) classical optimization problem: given a set of local longitudinal fields ($h_i$) and an interaction matrix ($J_{ij}$), find the assignment $\bf{s} = s_1 s_2 s_3 ... s_N$ , that minimizes the objective function $E(\bf{s})$, where,
\begin{equation}
E\left(\bf{s}\right) = \sum_{1 \leq i \leq N} h_i s_i + \sum_{1 \leq i \leq j \leq N} J_{ij}s_i s_j
\end{equation}
and $s_i \in {-1,1}$. Thus, the solution to this problem, $\bf{s}$, can be encoded into the ground-state wavefunction of the quantum Hamiltonian,
\begin{equation}
{\cal H}_p = \sum_{1 \leq i \leq N} h_i \sigma_i^z + \sum_{1 \leq i \leq j \leq N} J_{ij}\sigma_i^z \sigma_j^z.
\end{equation}

Quantum annealing exploits the adiabatic theorem of quantum mechanics, which states that a quantum system initialized in the ground state of a time-dependent Hamiltonian remains in the instantaneous ground state, as long as it is driven sufficiently slowly. Since the ground state of ${\cal H}_p$ encodes the solution to the optimization problem, the idea behind quantum annealing is to adiabatically prepare this ground state by initializing the quantum system in some easy-to-prepare ground state, ${\cal H}_b$. In this case, ${\cal H}_b$ corresponds to a superposition of all states of the computational basis. The system is driven slowly to the problem Hamiltonian, ${\cal H}(\tau = 1) \approx {\cal H}_p$. Deviations from the ground-state are expected due to deviations from adiabaticity, as well as thermal noise and imperfections in the implementation of the Hamiltonian.

Using the encoding methods discussed here, the authors were able to encode and to solve the global minima solution for small tetrapeptide and hexapeptide chains under several experimental schemes involving 5 and 8 qubits for four-amino-acid sequence (Hydrophobic-Polar model) and 5, 27, 28, and 81 qubits experiments for the six-amino-acid sequence under the Miyazawa-Jernigan model for general pairwise interactions.

\subsection{Six unit Miyazawa-Jernigan protein}
\begin{figure}[H]
 \centering
  \includegraphics[scale=0.2]{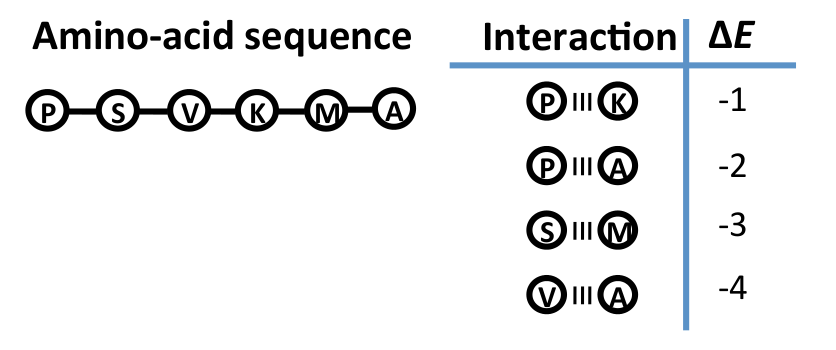} 
\caption{Interaction matrix for our protein in the MJ model.}
  \label{mj}
\end{figure}

The example we will present here is a different encoding of the largest problem performed in \cite{Perdomo-Ortiz2012}: the Miyazawa-Jernigan (MJ) protein, Proline-Serine-Valine-Lysine-Methionine-Alanine (PSVKMA) on a $2D$ lattice. We will use the pair-wise nearest-neighbor MJ interaction energies presented in Table 3 of \cite{Miyazawa1996} and shown in Fig.~\ref{mj}.
\noindent We will use the turn ancilla construction for our energy function and constrain the first three virtual bits to 010, as before. Recall that the turn ancilla construction requires $2N-5$ physical information bits; thus, our 6-unit MJ protein will be encoded into 7 bits.

\subsubsection{$E_{back}(\boldsymbol{q})$ for 6-unit SAW on $2D$ lattice}
Using Eq.~\ref{eback}, we find that our 6-unit protein has the backwards energy function,
\begin{align}
& E_{back}(\boldsymbol{q}) = \lambda_{back} ( q_1 q_2-2 q_1 q_3 q_2+2 q_3 q_2-2 q_3 q_4 q_2-2 q_3 q_5 q_2\\
& +4 q_3 q_4 q_5 q_2 -2 q_4 q_5 q_2+q_5 q_2 + q_3 q_4-2 q_3 q_4 q_5+2 q_4 q_5-2q_4 q_5 q_6\nonumber\\
& +q_5 q_6 +q_4 q_7-2 q_4 q_5 q_7-2 q_4 q_6 q_7+4 q_4 q_5 q_6 q_7-2 q_5 q_6 q_7+q_6 q_7 ).\nonumber
\end{align}
Soon, we will discuss how to choose the appropriate value for $\lambda_{back}$ but for now we simply note that $\lambda_{back}$ and $\lambda_{overlap}$ penalize the same illegal folds; thus we realize that $\lambda_{back} = \lambda_{overlap}$.

\subsubsection{$E_{overlap}(\boldsymbol{q})$ for 6-unit SAW on $2D$ lattice}
Using Eq.~\ref{overlapeq}, we calculate the overlap energy function as,
\small
\begin{align}
& E_{overlap}(\boldsymbol{q}) = \lambda_{overlap} (96 q_2 q_1-96 q_2 q_3 q_1-64 q_3 q_1-64 q_2 q_4 q_1+64 q_2 q_3 q_4 q_1-96 q_3 q_4 q_1+96 q_4 q_1\nonumber\\
& -96 q_2 q_5 q_1+64 q_2 q_4 q_5 q_1-96 q_4 q_5 q_1-64 q_5 q_1-48 q_2 q_6 q_1+32 q_2 q_3 q_6 q_1-48 q_3 q_6 q_1+32 q_3 q_4 q_6 q_1\nonumber\\
& -48 q_4 q_6 q_1+32 q_2 q_5 q_6 q_1+32 q_4 q_5 q_6 q_1-48 q_5 q_6 q_1+72 q_6 q_1-48 q_2 q_7 q_1-48 q_3 q_7 q_1+32 q_2 q_4 q_7 q_1\nonumber\\
& -48 q_4 q_7 q_1+96 q_3 q_5 q_7 q_1-48q_5 q_7 q_1+32 q_2 q_6 q_7 q_1+32 q_4 q_6 q_7 q_1-48 q_6 q_7 q_1-8 q_7 q_1-8 q_3 q_{10}\nonumber\\
& +64 q_3 q_8 q_1+64 q_5 q_8 q_1-32 q_8 q_1+32 q_3 q_9 q_1+32 q_5 q_9 q_1-16 q_9 q_1+16 q_3 q_{10} q_1+16 q_5 q_{10} q_1-8 q_{10} q_1\nonumber\\
& +8 q_3 q_{11} q_1+8 q_5 q_{11} q_1-4 q_{11} q_1+64 q_3 q_{12} q_1+64 q_5 q_{12} q_1+64 q_7 q_{12} q_1-96 q_{12} q_1+32 q_3 q_{13} q_1\nonumber\\
& +32 q_5 q_{13} q_1+32 q_7 q_{13} q_1-48 q_{13} q_1+16 q_3 q_{14} q_1+16 q_5 q_{14} q_1+16 q_7 q_{14} q_1-24 q_{14} q_1+8 q_3 q_{15} q_1\nonumber\\
& +8 q_5 q_{15} q_1+8 q_7 q_{15} q_1-12q_{15} q_1+64 q_1+144 q_2+96 q_2 q_3+64 q_3-64 q_2 q_4-64 q_2 q_3 q_4+96 q_3 q_4+144 q_4\nonumber\\
& +96 q_2 q_5-96 q_2 q_3 q_5-64 q_3 q_5-64 q_2 q_4 q_5+64 q_2 q_3 q_4 q_5-96 q_3 q_4 q_5+96 q_4 q_5+64 q_5-8 q_2 q_6-48 q_2 q_3 q_6\nonumber\\
& +72 q_3 q_6-48 q_2 q_4 q_6-48 q_3 q_4 q_6-8 q_4 q_6-48 q_2 q_5 q_6+32 q_2 q_3 q_5 q_6-48 q_3 q_5 q_6+32 q_3 q_4 q_5 q_6-48 q_4 q_5 q_6\nonumber\\
& +72 q_5 q_6+36 q_6+72 q_2 q_7-48 q_2 q_3 q_7-8 q_3 q_7-48 q_2 q_4 q_7+32 q_2 q_3 q_4 q_7-48 q_3 q_4 q_7+72 q_4 q_7-48 q_2 q_5 q_7\nonumber\\
& -48 q_3 q_5 q_7+32 q_2 q_4 q_5 q_7-48 q_4 q_5 q_7-8 q_5 q_7-48 q_2 q_6 q_7+32 q_2 q_3 q_6 q_7-48 q_3 q_6 q_7+32 q_3 q_4 q_6 q_7\nonumber\\
& +32 q_2 q_5 q_6 q_7+32 q_4 q_5 q_6 q_7-48 q_5 q_6 q_7+72 q_6 q_7+36 q_7-96 q_2 q_8-32 q_3 q_8+64 q_2 q_4 q_8-96 q_4 q_8\nonumber\\
& -32 q_5 q_8-32 q_8-48 q_2 q_9-16 q_3 q_9+32 q_2 q_4 q_9-48 q_4 q_9+32 q_3 q_5 q_9-16 q_5 q_9+64 q_8 q_9-32 q_9-24 q_2 q_{10}\nonumber\\
& +16 q_2 q_4 q_{10}-24 q_4 q_{10}+16 q_3 q_5 q_{10}-8 q_5 q_{10}+32 q_8 q_{10}+16 q_9 q_{10}-20 q_{10}-12 q_2 q_{11}-4 q_3 q_{11}+8 q_2 q_4 q_{11}\nonumber\\
& -12 q_4 q_{11}+8 q_3 q_5 q_{11}-4 q_5 q_{11}+16 q_8 q_{11}+8 q_9 q_{11}+4 q_{10} q_{11}-11 q_{11}-96 q_2 q_{12}-96 q_3 q_{12}+64 q_2 q_4 q_{12}\nonumber\\
& -96 q_4 q_{12}+64 q_3 q_5 q_{12}-96 q_5 q_{12}+64 q_2 q_6 q_{12}+64 q_4 q_6 q_{12}-96 q_6 q_{12}+64 q_3 q_7 q_{12}+64 q_5 q_7 q_{12}\nonumber\\
& -96 q_7 q_{12}+64 q_{12}-48 q_2 q_{13}-48 q_3 q_{13}+32 q_2 q_4 q_{13}-48 q_4 q_{13}+32 q_3 q_5 q_{13}-48 q_5 q_{13}+32 q_2 q_6 q_{13}\nonumber\\
& +32 q_4 q_6 q_{13}-48 q_6 q_{13}+32 q_3 q_7 q_{13}+32 q_5 q_7 q_{13}-48 q_7 q_{13}+64 q_{12} q_{13}+16 q_{13}-24 q_2 q_{14}-24 q_3 q_{14}\nonumber\\
& +16 q_2 q_4 q_{14}-24 q_4 q_{14}+16 q_3 q_5 q_{14}-24 q_5 q_{14}+16 q_2 q_6 q_{14}+16 q_4 q_6 q_{14}-24 q_6 q_{14}+16 q_3 q_7 q_{14}\nonumber\\
& +16 q_5 q_7 q_{14}-24 q_7 q_{14}+32 q_{12} q_{14}+16 q_{13}q_{14}+4 q_{14}-12 q_2 q_{15}-12 q_3 q_{15}+8 q_2 q_4 q_{15}-12 q_4 q_{15}\nonumber\\
& +8 q_3 q_5 q_{15}-12 q_5 q_{15}+8 q_2 q_6 q_{15}+8 q_4 q_6 q_{15}-12 q_6 q_{15}+8 q_3 q_7 q_{15}+8 q_5 q_7 q_{15}-12 q_7 q_{15}\nonumber\\
& +16 q_{12} q_{15}+8 q_{13} q_{15}+4 q_{14} q_{15}+q_{15}-48 q_4 q_6 q_7+64 q_3 q_5 q_8).
\end{align}
\normalsize
We notice that as discussed in Sec.~\ref{reduction}, all the 3-local terms here contain at least two physical information qubits (i.e. $q_1$ through $q_7$).

\subsubsection{$E_{pair}(\boldsymbol{q})$ for MJ-model PSVKMA}
Using the $J$ matrix as defined in Eq.~\ref{paireq} we calculate the pair-wise energy function as,
\small
\begin{align}
& E_{pair}(\boldsymbol{q}) = -4 q_2 q_{16}+4 q_1 q_3 q_{16}+3 q_{16}-8 q_1 q_{17}-16 q_2 q_{17}+8 q_1 q_3 q_{17}-8 q_3 q_{17}\nonumber\\
& +8 q_2 q_4 q_{17}-16 q_4 q_{17}+8 q_1 q_5 q_{17}+8 q_3 q_5 q_{17}-8 q_5 q_{17}+8 q_2 q_6 q_{17}+8 q_4 q_6 q_{17}-16 q_6 q_{17}\nonumber\\
& +8 q_1 q_7 q_{17}+8 q_3 q_7 q_{17}+8 q_5 q_7 q_{17}-8 q_7 q_{17}+30 q_{17}-12 q_1 q_{18}-12 q_2 q_{18}+12 q_1 q_3 q_{18}\nonumber\\
& -12 q_3 q_{18}+12 q_2 q_4 q_{18}-12 q_4 q_{18}+12 q_1 q_5 q_{18}+12 q_3 q_5 q_{18}-12 q_5 q_{18}+21 q_{18}-16 q_2 q_{19}\nonumber\\
& -16 q_3 q_{19}+16q_2 q_4 q_{19}-16 q_4 q_{19}+16 q_3 q_5 q_{19}-16 q_5 q_{19}+16 q_2 q_6 q_{19}+16 q_4 q_6 q_{19}\nonumber\\
& -16 q_6 q_{19} +16 q_3 q_7 q_{19}+16 q_5 q_7 q_{19}-16 q_7 q_{19}+28 q_{19}.
\end{align}
\normalsize

\subsubsection{Setting $\lambda$ penalty values}
Finally, we will discuss how one chooses the correct penalty values for the energy function. This is a crucial step if one wishes to implement the algorithm experimentally as all currently available architectures for adiabatic quantum annealing have limited coupler resolution. That is, quantum annealing machines cannot realize arbitrary constant values for the QUBO expression. Thus, it is very important that one chooses the lowest possible penalty values which still impose the correct constraints. In our problem we choose the value of $\lambda_{overlap}$ by asking ourselves: what is the greatest possible amount that any overlap could \emph{lower} the system energy? In general, a very conservative upper bound can be obtained by simply summing together every $J$ matrix element (which would mean that a single overlap allowed every single possible interaction to occur); in our problem this upper-bound would be -10. Thus, we can set $\lambda_{overlap} = +10$.

\subsubsection{Reduction to 2-local}
Using a standard greedy search algorithm we find that an efficient way to collapse this energy function to 2-local is to make ancilla with the qubit pairs, 
\begin{equation}
\begin{aligned}
q_2 q_4 & \rightarrow q_{20} &\\ 
q_1 q_3 & \rightarrow q_{21} &\\
q_3 q_5 & \rightarrow q_{22} &\\
q_1 q_5 & \rightarrow q_{23} &\\
q_2 q_6 & \rightarrow q_{24} &\\
q_4 q_6 & \rightarrow q_{25} &\\
q_3 q_7 & \rightarrow q_{26} &\\
q_5 q_7 & \rightarrow q_{27} &\\
q_1 q_7 & \rightarrow q_{28} &.
\end{aligned}
\label{delta}
\end{equation}
There is one issue left to discuss - the value of $\delta_n$ in Eq.~\ref{reduce}. The purpose of $\delta_n$ is to constrain the reductions in Eq.~\ref{delta} so that the value of the ancillary bit actually corresponds to the product of the two bits it is supposed to represent. To understand how Eq.~\ref{reduce} accomplishes this see Table~\ref{truthtable}. In order for Eq.~\ref{reduce} to work we must choose $\delta_n$ which is large enough so that a violation of the reduction we desire will always raise the system energy. Thus, we must ensure that $\delta_n$ is large enough so that configurations which do not conform to the reduction are penalized by an amount higher than the largest penalty they could avoid and larger in magnitude than the largest energy reduction they could achieve with the illegal move. However, finding the exact minimum value of $E(\boldsymbol{q})$ can be an extremely difficult problem. Instead, we can simply make an upper-bound for the penalty by setting it equal to one plus either the sum of the absolute value of all pseudo-boolean coefficients corresponding to the variables being collapsed in $E(\boldsymbol{q})$ \cite{Babbush2013b}.

\begin{table}
\label{truthtable}
\centering
\caption{Truth table for the function $E_\wedge (q_i, q_j, \tilde q_n;\delta_n)$ from Eq.~\ref{reduce}.}
\begin{tabular}{c c c c}
\hline
\hline
$q_n$ & $q_i$ & $q_j$ & $E_\wedge (q_i, q_j, \tilde q_n;\delta_n)$ \\
\hline
0 & 0 & 0 & 0  \\
0 & 0 & 1 & 0  \\
0 & 1 & 0 & 0  \\
1 & 1 & 1 & 0  \\
1 & 0 & 0 & $3 \delta$  \\
1 & 0 & 1 & $\delta$  \\
1 & 1 & 0 & $\delta$  \\
0 & 1 & 1 & $\delta$  \\
\hline
\hline
\end{tabular}
\end{table}

\subsubsection{QUBO Matrix and Solutions}
After reduction of the energy function to 2-local, we arrive at the final pseudo-boolean energy function. Instead of writing out the entire pseudo-boolean expression we will instead provide a matrix containing all of the coefficients of 1-local terms on the diagonal and 2-local terms in the upper triangular portion of this matrix. This representation is known as the QUBO matrix and contains all of the couplings needed for experimental implementation and is shown in Eq.~\ref{qmat}. Note that the full pseudo-boolean expression contains one constant term that we drop in the matrix representation. This constant has a value of $C = 180$ for this particular problem.
{\setlength{\arraycolsep}{1pt}
\begin{equation}
\label{qmat}
\scalemath{0.45}{
\left(
\begin{array}{cccccccccccccccccccccccccccc}
 320 & 485 & 42962 & 480 & 42962 & 360 & 42962 & -160 & -80 & -40 & -20 & -480 & -240 & -120 & -60 & 0 & -8 & -12 & 0 & -320 & -85924 & 0 & -85924 & -240 & -240 & 0 & 0 & -85924 \\
 0 & 720 & 490 & 42962 & 485 & 42962 & 360 & -480 & -240 & -120 & -60 & -480 & -240 & -120 & -60 & -4 & -16 & -12 & -16 & -85924 & -490 & -490 & -480 & -85924 & 0 & -240 & -240 & -240 \\
 0 & 0 & 320 & 485 & 42962 & 360 & 42962 & -160 & -80 & -40 & -20 & -480 & -240 & -120 & -60 & 0 & -8 & -12 & -16 & -330 & -85924 & -85924 & 0 & -240 & -240 & -85924 & 0 & 0 \\
 0 & 0 & 0 & 720 & 490 & 42962 & 365 & -480 & -240 & -120 & -60 & -480 & -240 & -120 & -60 & 0 & -16 & -12 & -16 & -85924 & -480 & -490 & -480 & 0 & -85924 & -240 & -250 & -240 \\
 0 & 0 & 0 & 0 & 320 & 365 & 42962 & -160 & -80 & -40 & -20 & -480 & -240 & -120 & -60 & 0 & -8 & -12 & -16 & -330 & 0 & -85924 & -85924 & -240 & -250 & 0 & -85924 & 0 \\
 0 & 0 & 0 & 0 & 0 & 180 & 365 & 0 & 0 & 0 & 0 & -480 & -240 & -120 & -60 & 0 & -16 & 0 & -16 & -240 & -240 & -240 & -240 & -85924 & -85924 & -240 & -250 & -240 \\
 0 & 0 & 0 & 0 & 0 & 0 & 180 & 0 & 0 & 0 & 0 & -480 & -240 & -120 & -60 & 0 & -8 & 0 & -16 & -240 & -240 & -240 & -240 & -240 & -250 & -85924 & -85924 & -85924 \\
 0 & 0 & 0 & 0 & 0 & 0 & 0 & -160 & 320 & 160 & 80 & 0 & 0 & 0 & 0 & 0 & 0 & 0 & 0 & 320 & 320 & 320 & 320 & 0 & 0 & 0 & 0 & 0 \\
 0 & 0 & 0 & 0 & 0 & 0 & 0 & 0 & -160 & 80 & 40 & 0 & 0 & 0 & 0 & 0 & 0 & 0 & 0 & 160 & 160 & 160 & 160 & 0 & 0 & 0 & 0 & 0 \\
 0 & 0 & 0 & 0 & 0 & 0 & 0 & 0 & 0 & -100 & 20 & 0 & 0 & 0 & 0 & 0 & 0 & 0 & 0 & 80 & 80 & 80 & 80 & 0 & 0 & 0 & 0 & 0 \\
 0 & 0 & 0 & 0 & 0 & 0 & 0 & 0 & 0 & 0 & -55 & 0 & 0 & 0 & 0 & 0 & 0 & 0 & 0 & 40 & 40 & 40 & 40 & 0 & 0 & 0 & 0 & 0 \\
 0 & 0 & 0 & 0 & 0 & 0 & 0 & 0 & 0 & 0 & 0 & 320 & 320 & 160 & 80 & 0 & 0 & 0 & 0 & 320 & 320 & 320 & 320 & 320 & 320 & 320 & 320 & 320 \\
 0 & 0 & 0 & 0 & 0 & 0 & 0 & 0 & 0 & 0 & 0 & 0 & 80 & 80 & 40 & 0 & 0 & 0 & 0 & 160 & 160 & 160 & 160 & 160 & 160 & 160 & 160 & 160 \\
 0 & 0 & 0 & 0 & 0 & 0 & 0 & 0 & 0 & 0 & 0 & 0 & 0 & 20 & 20 & 0 & 0 & 0 & 0 & 80 & 80 & 80 & 80 & 80 & 80 & 80 & 80 & 80 \\
 0 & 0 & 0 & 0 & 0 & 0 & 0 & 0 & 0 & 0 & 0 & 0 & 0 & 0 & 5 & 0 & 0 & 0 & 0 & 40 & 40 & 40 & 40 & 40 & 40 & 40 & 40 & 40 \\
 0 & 0 & 0 & 0 & 0 & 0 & 0 & 0 & 0 & 0 & 0 & 0 & 0 & 0 & 0 & 3 & 0 & 0 & 0 & 0 & 4 & 0 & 0 & 0 & 0 & 0 & 0 & 0 \\
 0 & 0 & 0 & 0 & 0 & 0 & 0 & 0 & 0 & 0 & 0 & 0 & 0 & 0 & 0 & 0 & 30 & 0 & 0 & 8 & 8 & 8 & 8 & 8 & 8 & 8 & 8 & 8 \\
 0 & 0 & 0 & 0 & 0 & 0 & 0 & 0 & 0 & 0 & 0 & 0 & 0 & 0 & 0 & 0 & 0 & 21 & 0 & 12 & 12 & 12 & 12 & 0 & 0 & 0 & 0 & 0 \\
 0 & 0 & 0 & 0 & 0 & 0 & 0 & 0 & 0 & 0 & 0 & 0 & 0 & 0 & 0 & 0 & 0 & 0 & 28 & 16 & 0 & 16 & 0 & 16 & 16 & 16 & 16 & 0 \\
 0 & 0 & 0 & 0 & 0 & 0 & 0 & 0 & 0 & 0 & 0 & 0 & 0 & 0 & 0 & 0 & 0 & 0 & 0 & 128566 & 320 & 340 & 320 & 0 & 0 & 160 & 160 & 160 \\
 0 & 0 & 0 & 0 & 0 & 0 & 0 & 0 & 0 & 0 & 0 & 0 & 0 & 0 & 0 & 0 & 0 & 0 & 0 & 0 & 128566 & 0 & 0 & 160 & 160 & 0 & 480 & 0 \\
 0 & 0 & 0 & 0 & 0 & 0 & 0 & 0 & 0 & 0 & 0 & 0 & 0 & 0 & 0 & 0 & 0 & 0 & 0 & 0 & 0 & 128566 & 0 & 160 & 160 & 0 & 0 & 0 \\
 0 & 0 & 0 & 0 & 0 & 0 & 0 & 0 & 0 & 0 & 0 & 0 & 0 & 0 & 0 & 0 & 0 & 0 & 0 & 0 & 0 & 0 & 128566 & 160 & 160 & 0 & 0 & 0 \\
 0 & 0 & 0 & 0 & 0 & 0 & 0 & 0 & 0 & 0 & 0 & 0 & 0 & 0 & 0 & 0 & 0 & 0 & 0 & 0 & 0 & 0 & 0 & 128846 & 0 & 160 & 160 & 160 \\
 0 & 0 & 0 & 0 & 0 & 0 & 0 & 0 & 0 & 0 & 0 & 0 & 0 & 0 & 0 & 0 & 0 & 0 & 0 & 0 & 0 & 0 & 0 & 0 & 128846 & 160 & 180 & 160 \\
 0 & 0 & 0 & 0 & 0 & 0 & 0 & 0 & 0 & 0 & 0 & 0 & 0 & 0 & 0 & 0 & 0 & 0 & 0 & 0 & 0 & 0 & 0 & 0 & 0 & 128846 & 0 & 0 \\
 0 & 0 & 0 & 0 & 0 & 0 & 0 & 0 & 0 & 0 & 0 & 0 & 0 & 0 & 0 & 0 & 0 & 0 & 0 & 0 & 0 & 0 & 0 & 0 & 0 & 0 & 128846 & 0 \\
 0 & 0 & 0 & 0 & 0 & 0 & 0 & 0 & 0 & 0 & 0 & 0 & 0 & 0 & 0 & 0 & 0 & 0 & 0 & 0 & 0 & 0 & 0 & 0 & 0 & 0 & 0 & 128846 \\
\end{array}
\right)
}
\end{equation}
}

Taking the matrix in Eq.~\ref{qmat} as $Q$, we can write the total energy of a given solution (denoted by $\boldsymbol{q}$) as,
\begin{equation}
E(\boldsymbol{q}) = \boldsymbol{q} Q \boldsymbol{q}.
\label{qubo}
\end{equation}
The problem is now ready for its implementation on a quantum device. For our particular problem instance the solution string is given by the bit string,
\begin{equation}
0, 0, 0, 1, 0, 1, 1, 1, 1, 0, 0, 1, 1, 1, 0, 0, 1, 0, 1, 0, 0, 0, 0, 0, 1, 0, 0, 0.
\end{equation}
The energy given by Eq.~\ref{qubo} is $-186$. In the original expression this corresponds to an energy of $C - 186 = 180 - 186 = -6$. Let's confirm that this is accurate to the MJ model. Looking only at the physical information bits and prepending the first three constant bits ($010$) we see that the bit string prescribes the following fold:
\begin{equation}
\boldsymbol{q} = \underbrace{0 1}_{\text{right}}\underbrace{0 0}_{\text{down}}\underbrace{0 0}_{\text{down}}\underbrace{1 0}_{\text{left}}\underbrace{1 1}_{\text{up}}
\end{equation}
which corresponds to the fold,
\begin{figure}[h]
 \centering
  \includegraphics[scale=0.5]{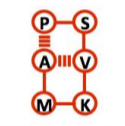}
 \caption{The solution to our example problem for MJ protein PSVKMA.}
 \label{graph}
\end{figure}

\addtocontents{toc}{\protect\enlargethispage{2\baselineskip}}
\newpage
\section{Conclusion}

As both traditional and quantum computer science continue to advance as fields, domain scientists from all disciplines need to develop new ways of representing problems in order to leverage state-of-the-art computational tools. In this review, we discussed strategies and techniques for solving lattice heteropolymer problems with some of these tools. While the lattice heteropolymer model is widely applicable to many problems, the general principles used to optimally encode and constrain this particular application are fairly universal for discrete optimization problems in the physical sciences.

We focused on three mappings: ``turn ancilla'', ``turn circuit'' and ``diamond''. The turn ancilla mapping is the best mapping in terms of the scaling of the number of resources for large instances, thus making it ideal for benchmark studies of lattice folding using (heuristic) solvers for pseudo-boolean minimization. Additionally, this method shows how one can use ancilla variables to construct a fitness function with relatively few constraints per clause (i.e. low-locality). With ancilla variables even an extremely simple encoding, such as the turn encoding, can be used to construct a complicated energy function. While some of the particular tricks employed to optimize the efficiency of this mapping, such as introducing the backwards penalty, are specific to lattice heteropolymers, the general logic behind these tricks is much more universal.

The turn circuit mapping is the most compact of all three mappings. The extremely efficient use of variables (qubits) makes it ideal for benchmark experiments on quantum devices which can handle many body couplings. Moreover, the turn circuit method demonstrates how one can construct an elaborate energy function by utilizing logic circuits to put together a high-local fitness function of arbitrary complexity without ancilla variables. While different problems may involve different circuits, the underlying strategy is very broadly applicable.

The diamond encoding illustrates a strategy for producing an extremely under-constrained optimization problem. Furthermore, this method demonstrates that even fairly complex energy functions can be represented as natively 2-local functions if one is willing to sacrifice efficiency. Many quantum devices can only couple bits pairwise; thus, this is a very important quality of the diamond encoding. Finally, if one uses another, more efficient encoding, we explain how reductions can be used to replace high-local terms with 2-local terms in an optimally efficient fashion but at the cost of needing very high coupler resolution. The relatively few constraints in the diamond encoding make it a natural choice for exact or heuristic ILP and W-SAT solvers.

These three strategies elucidate many of the concepts that we find important when producing problems suitable for the D-Wave device utilized in \cite{Perdomo-Ortiz2012}. Accordingly, as quantum information science continues to develop, we hope that the methods discussed in this review will be useful to scientists wishing to leverage similar technology for the solution of discrete optimization problems.

\section{Acknowledgements}

The authors thank Joseph Goodknight for help editing this chapter. This research was sponsored by the United States Department of Defense. The views and conclusions contained in this document are those of the authors and should not be interpreted as representing the official policies, either expressly or implied, of the U.S. Government.

\bibliographystyle{ieeetr}
\bibliography{./library}

\end{document}